\documentclass{aa}  

\usepackage{graphicx}
\usepackage{amsmath,amsfonts,amssymb}
\usepackage{natbib}

\usepackage{txfonts}
\usepackage{xcolor}

\usepackage{blindtext}
\usepackage{float}
\usepackage{dblfloatfix}
\usepackage{afterpage}
\usepackage{ifthen}
\usepackage[morefloats=12]{morefloats}
\usepackage{threeparttable}

\usepackage{placeins}
\usepackage{multicol}
\usepackage[switch]{lineno}
\definecolor{linkcolor}{rgb}{0.6,0,0}
\definecolor{citecolor}{rgb}{0,0,0.75}
\definecolor{urlcolor}{rgb}{0.12,0.46,0.7}
\usepackage[breaklinks, colorlinks, urlcolor=urlcolor,
    linkcolor=linkcolor,citecolor=citecolor,pdfencoding=auto]{hyperref}
\hypersetup{linktocpage}
\usepackage{bold-extra}

\usepackage[nameinlink,capitalise]{cleveref}
\crefname{section}{Sect.}{Sects.}
\crefname{table}{Table}{Tables}
\crefname{equation}{Eq.}{Eqs.}
\crefname{appendix}{Appendix}{Appendices}
\Crefname{section}{Section}{Sections}
\Crefname{table}{Table}{Tables}
\Crefname{equation}{Equation}{Equations}
\crefname{appendix}{Appendix}{Appendices}
\makeatletter
\AddToHook{cmd/appendix/before}{\def\cref@section@alias{appendix}}
\makeatother

\usepackage{caption}
\usepackage{subcaption}

\DeclareRobustCommand{\ion}[2]{%
\relax\ifmmode
\ifx\testbx\f@series
{\mathbf{#1\,\mathsc{#2}}}\else
{\mathrm{#1\,\mathsc{#2}}}\fi
\else\textup{#1\,{\mdseries\textsc{#2}}}%
\fi}

\def\setsymbol#1#2{\expandafter\def\csname #1\endcsname{#2}}
\def\getsymbol#1{\csname #1\endcsname}

\def\Planck{\textit{Planck}}

\newbox\tablebox    \newdimen\tablewidth
\def\leaderfil{\leaders\hbox to 5pt{\hss.\hss}\hfil}
\def\endPlancktable{\tablewidth=\columnwidth 
    $$\hss\copy\tablebox\hss$$
    \vskip-\lastskip\vskip -2pt}

\def\tablenote#1 #2\par{\begingroup \parindent=0.8em
    \abovedisplayshortskip=0pt\belowdisplayshortskip=0pt
    \noindent
    $$\hss\vbox{\hsize\tablewidth \hangindent=\parindent \hangafter=1 \noindent
    \hbox to \parindent{$^#1$\hss}\strut#2\strut\par}\hss$$
    \endgroup}
\def\doubleline{\vskip 3pt\hrule \vskip 1.5pt \hrule \vskip 5pt}

\def\L2{\ifmmode L_2\else $L_2$\fi}

\def\DeltaT{\ifmmode \Delta T\else $\Delta T$\fi}
\def\deltat{\ifmmode \Delta t\else $\Delta t$\fi}
\def\fknee{\ifmmode f_{\rm knee}\else $f_{\rm knee}$\fi}
\def\Fmax{\ifmmode F_{\rm max}\else $F_{\rm max}$\fi}
\def\solar{\ifmmode{\rm M}_{\mathord\odot}\else${\rm M}_{\mathord\odot}$\fi}
\def\Msolar{\ifmmode{\rm M}_{\mathord\odot}\else${\rm M}_{\mathord\odot}$\fi}
\def\Lsolar{\ifmmode{\rm L}_{\mathord\odot}\else${\rm L}_{\mathord\odot}$\fi}
\def\inv{\ifmmode^{-1}\else$^{-1}$\fi}
\def\mo{\ifmmode^{-1}\else$^{-1}$\fi}
\def\sup#1{\ifmmode ^{\rm #1}\else $^{\rm #1}$\fi}
\def\expo#1{\ifmmode \times 10^{#1}\else $\times 10^{#1}$\fi}
\def\,{\thinspace}
\def\lsim{\mathrel{\raise .4ex\hbox{\rlap{$<$}\lower 1.2ex\hbox{$\sim$}}}}
\def\gsim{\mathrel{\raise .4ex\hbox{\rlap{$>$}\lower 1.2ex\hbox{$\sim$}}}}

\def\simprop{\mathrel{\raise .4ex\hbox{\rlap{$\propto$}\lower 1.2ex\hbox{$\sim$}}}}
\def\deg{\ifmmode^\circ\else$^\circ$\fi}
\def\pdeg{\ifmmode $\setbox0=\hbox{$^{\circ}$}\rlap{\hskip.11\wd0 .}$^{\circ}
          \else \setbox0=\hbox{$^{\circ}$}\rlap{\hskip.11\wd0 .}$^{\circ}$\fi}
\def\arcs{\ifmmode {^{\scriptstyle\prime\prime}}
          \else $^{\scriptstyle\prime\prime}$\fi}
\def\arcm{\ifmmode {^{\scriptstyle\prime}}
          \else $^{\scriptstyle\prime}$\fi}
\newdimen\sa  \newdimen\sb
\def\parcs{\sa=.07em \sb=.03em
     \ifmmode \hbox{\rlap{.}}^{\scriptstyle\prime\kern -\sb\prime}\hbox{\kern -\sa}
     \else \rlap{.}$^{\scriptstyle\prime\kern -\sb\prime}$\kern -\sa\fi}
\def\parcm{\sa=.08em \sb=.03em
     \ifmmode \hbox{\rlap{.}\kern\sa}^{\scriptstyle\prime}\hbox{\kern-\sb}
     \else \rlap{.}\kern\sa$^{\scriptstyle\prime}$\kern-\sb\fi}
\def\ra[#1 #2 #3.#4]{#1\sup{h}#2\sup{m}#3\sup{s}\llap.#4}
\def\dec[#1 #2 #3.#4]{#1\deg#2\arcm#3\arcs\llap.#4}
\def\deco[#1 #2 #3]{#1\deg#2\arcm#3\arcs}
\def\rra[#1 #2]{#1\sup{h}#2\sup{m}}

\def\dots{\relax\ifmmode \ldots\else $\ldots$\fi}
\def\WHzsr{\ifmmode $W\,Hz\mo\,sr\mo$\else W\,Hz\mo\,sr\mo\fi}
\def\mHz{\ifmmode $\,mHz$\else \,mHz\fi}
\def\GHz{\ifmmode $\,GHz$\else \,GHz\fi}
\def\mKs{\ifmmode $\,mK\,s$^{1/2}\else \,mK\,s$^{1/2}$\fi}
\def\muKs{\ifmmode \,\mu$K\,s$^{1/2}\else \,$\mu$K\,s$^{1/2}$\fi}
\def\muKRJs{\ifmmode \,\mu$K$_{\rm RJ}$\,s$^{1/2}\else \,$\mu$K$_{\rm RJ}$\,s$^{1/2}$\fi}
\def\muKHz{\ifmmode \,\mu$K\,Hz$^{-1/2}\else \,$\mu$K\,Hz$^{-1/2}$\fi}
\def\MJysr{\ifmmode \,$MJy\,sr\mo$\else \,MJy\,sr\mo\fi}
\def\MJysrmK{\ifmmode \,$MJy\,sr\mo$\,mK$_{\rm CMB}\mo\else \,MJy\,sr\mo\,mK$_{\rm CMB}\mo$\fi}
\def\microns{\ifmmode \,\mu$m$\else \,$\mu$m\fi}

\def\muK{\ifmmode \,\mu$K$\else \,$\mu$\hbox{K}\fi}
\def\microK{\ifmmode \,\mu$K$\else \,$\mu$\hbox{K}\fi}
\def\muW{\ifmmode \,\mu$W$\else \,$\mu$\hbox{W}\fi}
\def\kms{\ifmmode $\,km\,s$^{-1}\else \,km\,s$^{-1}$\fi}
\def\kmsMpc{\ifmmode $\,\kms\,Mpc\mo$\else \,\kms\,Mpc\mo\fi}

\providecommand{\sorthelp}[1]{}

\newcommand{\mathsc}[1]{{\normalfont\textsc{#1}}}

\def\Cosmoglobe{\textsc{Cosmoglobe}}
\def\cosmoglobe{\textsc{Cosmoglobe}}
\def\BeyondPlanck{\textsc{BeyondPlanck}}
\def\Planck{\textit{Planck}}
\def\planck{\textit{Planck}}

\def\npipe{NPIPE}

\def\WMAP{\textit{WMAP}}
\def\COBE{\textit{COBE}}
\def\GAIA{\textit{Gaia}}

\def\Gaia{\textit{Gaia}}
\def\Ha{H$\alpha$}
\def\litebird{\textit{LiteBIRD}}

\def\nside{$N_{\mathrm{side}}$}

\def\Commander{\texttt{Commander} }

\def\Tcmb{\ifmmode T_\mathrm{CMB}\else $T_{\mathrm{CMB}}$\fi}
\def\Tcold{\ifmmode T_\mathrm{cold}\else $T_{\mathrm{cold}}$\fi}
\def\Thot{\ifmmode T_\mathrm{hot}\else $T_{\mathrm{hot}}$\fi}
\def\Tnear{\ifmmode T_\mathrm{near}\else $T_{\mathrm{near}}$\fi}
\def\scmb{\ifmmode s_\mathrm{CMB}\else $s_{\mathrm{CMB}}$\fi}
\def\squad{\ifmmode s_\mathrm{quad}\else $s_{\mathrm{quad}}$\fi}
\def\ssynch{\ifmmode s_\mathrm{s}\else $s_\mathrm{s}$\fi}
\def\sdust{\ifmmode s_\mathrm{d}\else $s_{\mathrm{d}}$\fi}
\def\ssdust{\ifmmode s_\mathrm{sd}\else $s_{\mathrm{sd}}$\fi}
\def\same{\ifmmode s_\mathrm{AME}\else $s_{\mathrm{AME}}$\fi}
\def\ssrc{\ifmmode s_\mathrm{src}\else $s_{\mathrm{src}}$\fi}
\def\sco{\ifmmode s_\mathrm{CO}\else $s_{\mathrm{CO}}$\fi}
\def\sff{\ifmmode s_\mathrm{ff}\else $s_{\mathrm{ff}}$\fi}
\def\gff{\ifmmode g_\mathrm{ff}\else $g_{\mathrm{ff}}$\fi}
\def\fsynch{\ifmmode f_\mathrm{s}\else $f_{\mathrm{s}}$\fi}
\def\fsd{\ifmmode f_\mathrm{sd}\else $f_{\mathrm{sd}}$\fi}
\def\fame{\ifmmode f_\mathrm{AME}\else $f_{\mathrm{AME}}$\fi}
\def\alphasrc{\ifmmode \alpha_\mathrm{src}\else $\alpha_{\mathrm{src}}$\fi}
\def\bcold{\ifmmode \beta_\mathrm{cold}\else $\beta_{\mathrm{cold}}$\fi}
\def\bhot{\ifmmode \beta_\mathrm{hot}\else $\beta_{\mathrm{hot}}$\fi}
\def\bnear{\ifmmode \beta_\mathrm{n}\else $\beta_{\mathrm{n}}$\fi}
\def\bsynch{\ifmmode \beta_\mathrm{s}\else $\beta_{\mathrm{s}}$\fi} 
\def\bsun{\ifmmode \beta_\mathrm{sun}\else $\beta_{\mathrm{sun}}$\fi} 
\def\nuzeros{\ifmmode \nu_{0,\mathrm{s}}\else $\nu_{0,\mathrm{s}}$\fi} 
\def\nuzeroff{\ifmmode \nu_{0,\mathrm{ff}}\else $\nu_{0,\mathrm{ff}}$\fi} 
\def\nuzerocold{\ifmmode \nu_{0,\mathrm{c}}\else $\nu_{0,\mathrm{c}}$\fi}
\def\nuzerohot{\ifmmode \nu_{0,\mathrm{h}}\else $\nu_{0,\mathrm{h}}$\fi}
\def\nuzeronear{\ifmmode \nu_{0,\mathrm{n}}\else $\nu_{0,\mathrm{n}}$\fi} 
\def\nuzeroame{\ifmmode \nu_{0,\mathrm{AME}}\else $\nu_{0,\mathrm{AME}}$\fi} 
\def\nuzerosd{\ifmmode \nu_{0,\mathrm{}}\else $\nu_{0,\mathrm{sd}}$\fi} 
\def\nuzerosrc{\ifmmode \nu_{0,\mathrm{src}}\else $\nu_{0,\mathrm{src}}$\fi} 
\def\nup{\ifmmode \nu_{\mathrm{p}}\else $\nu_{\mathrm{p}}$\fi} 
\def\alphasd{\ifmmode \alpha_{\mathrm{sd}}\else $\alpha_{\mathrm{sd}}$\fi} 
\def\Te{\ifmmode T_{\mathrm{e}}\else $T_{\mathrm{e}}$\fi} 
\def\kB{\ifmmode k_\mathrm{B}\else $k_{\mathrm{B}}$\fi}

\definecolor{darkpastelgreen}{rgb}{0.01, 0.75, 0.24}
\usepackage[normalem]{ulem} %

\begin{document} 

\title{\bfseries{\Cosmoglobe\ DR2. VI. Disentangling hot and cold thermal dust emission with Planck HFI}}

   \newcommand{\oslo}[0]{1}
\newcommand{\milan}[0]{2}
\newcommand{\ijclab}[0]{3}
\newcommand{\gothenberg}[0]{4}
\newcommand{\trento}[0]{5}
\newcommand{\milanoinfn}[0]{6}
\author{\small
R.~M.~Sullivan\inst{\oslo}\thanks{Corresponding author: R.~M.~Sullivan; \url{raelyn.sullivan@astro.uio.no}}
\and
E.~Gjerl\o w\inst{\oslo}
\and
M.~Galloway\inst{\oslo}
\and
D.~J.~Watts\inst{\oslo}
\and
R.~Aurvik\inst{\oslo}
\and
A.~Basyrov\inst{\oslo}
\and
L.~A.~Bianchi\inst{\oslo}
\and
A.~Bonato\inst{\milan}
\and
M.~Brilenkov\inst{\oslo}
\and
H.~K.~Eriksen\inst{\oslo}
\and
U.~Fuskeland\inst{\oslo}
\and
K.~A.~Glasscock\inst{\oslo}
\and
L.~T.~Hergt\inst{\ijclab}
\and
D.~Herman\inst{\oslo}
\and
J.~G.~S.~Lunde\inst{\oslo}
\and
A.~I.~Silva Martins\inst{\oslo}
\and
M.~San\inst{\oslo}
\and
D.~Sponseller\inst{\gothenberg}
\and
N.-O.~Stutzer\inst{\oslo}
\and
H.~Thommesen\inst{\oslo}
\and
V.~Vikenes\inst{\oslo}
\and
I.~K.~Wehus\inst{\oslo}
\and
L.~Zapelli\inst{\milan, \trento, \milanoinfn}
}
\institute{\small
Institute of Theoretical Astrophysics, University of Oslo, Blindern, Oslo, Norway\goodbreak
\and
Dipartimento di Fisica, Università degli Studi di Milano, Via Celoria, 16, Milano, Italy
\and
Laboratoire de Physique des 2 infinis -- Irène Joliot Curie (IJCLab), Orsay, France
\and
Department of Space, Earth and Environment, Chalmers University of Technology, Gothenburg, Sweden\goodbreak
\and
Università di Trento, Università degli Studi di Milano, CUP E66E23000110001\goodbreak
\and
INFN sezione di Milano, 20133 Milano, Italy\goodbreak
}

   \titlerunning{Disentangling cold and hot dust with HFI}
   \authorrunning{Sullivan et al.}

   \date{\today} 
   
   \abstract{
	   We present a four-component high-resolution model of thermal dust emission for microwave and sub-mm frequencies derived from \textit{Planck} HFI, WHAM and \textit{Gaia}. This model is inspired by a joint low-resolution template-based analysis of \textit{Planck} HFI and \textit{COBE}-DIRBE data presented in a companion paper, and the resulting high-resolution model derived here forms the basis for the thermal dust model employed in the \textsc{Cosmoglobe} DR2 reanalysis of \textit{COBE}-DIRBE. The four dust components are called ``cold dust'', ``hot dust'',  ``nearby dust'', and ``\Ha\ correlated dust'', respectively, and trace different physical environments. The spatial distributions of the nearby dust and \Ha\ dust components are defined by the Edenhofer et al.\ \textit{Gaia} 3D extinction model and the Wisconsin \Ha\ mapper (WHAM) survey, respectively, while the hot and cold dust components are fit freely pixel-by-pixel to the \textit{Planck} HFI data. 
       We use a global parameter grid search coupled to an amplitude map Gibbs sampler to fit this model to \textit{Planck} HFI data. In agreement with the companion low-resolution analysis, we find that the hot dust component is strongly correlated with the FIRAS \ion{C}{ii} map, while the cold dust component is strongly correlated with the HI4PI \ion{H}{i} map.
 Despite its fewer degrees of freedom per pixel compared to the \textit{Planck} 2015 legacy dust model, we find that this new model performs competitively in terms of overall residuals, capturing between 98 and 99.9\,\% of the full-sky dust variance for all channels between 100 and 857\,GHz. When fitting a spatially varying 3-parameter MBB model to the new four-component dust model with isotropic SEDs, we find very similar spatial distributions of $T$ and $\beta$ to those of the official \textit{Planck} analysis, and this new model thus represents an economical decomposition of previously published spatially varying spectral parameter maps. 
 We conclude that this new model represents both a statistically more efficient summary of thermal dust in the microwave and far-infrared regimes and a physically more realistic decomposition of the sky compared to the traditional 3-parameter MBB model.
 Finally, we expect that the novel templates of hot and cold dust emission presented in this paper may form a key component in future temperature and polarization measurements for projects such as Simons Observatory and \litebird, where high-precision measurements of dust will be critical for constraining $B$-mode CMB polarization. 
   }

   \maketitle

\section{Introduction}
Observing the Universe through a screen of foregrounds offers opportunities for both discovery and frustration \citep[e.g.,][]{Planck2018xii,planck2016-l04,bicep2_2014}. The foregrounds ultimately complicate our understanding of the backgrounds, and when poorly understood, lead to increased uncertainties or even misinterpretations. A better understanding of the foregrounds allows us to uncover the physics permeating matter on all scales, from within our solar system \citep[e.g.,][]{K98}, to the Milky Way \citep[e.g.,][]{planck2014-a12}, to distant galaxies \citep[e.g.,][]{planck2014-XXV,planck2014-XXVI,planck2011-5.2a}, while simultaneously allowing us to better observe the far away signals of the Cosmic Microwave Background (CMB; \citealp{penzias:1965}) and the Cosmic Infrared Background (CIB; \citep{partridge1967}), among others. Of particular interest to this paper and the cosmology community at large is the composition of thermal dust \citep[e.g.,][]{planck2013-p06b,Hensley2023}. Thermal dust represents a significant noise contribution to future measurements of primordial polarization $B$ modes \citep[][]{Bicep2018limit}, predicted to have been generated by gravitational waves during inflation, which is the primary goal of several ongoing or upcoming CMB experiments \citep{litebird2022,SO2019}. 

Aside from the aforementioned dust, the Milky Way contains gas (generating emission lines such as singly ionized carbon, \ion{C}{ii}; hydrogen, such as \Ha; neutral hydrogen, \ion{H}{i}; carbon monoxide, CO; and many more), stars, spinning dust, ions, electrons (interacting with magnetic fields to produce synchrotron and free-free radiation), and dark matter (only thus far interacting gravitationally). Each of these foregrounds trace relevant regions of the Milky Way and the interstellar medium (ISM). Since dust and gas live side-by-side, it is natural that one may trace the other, and indeed it is known that neutral hydrogen (as observed by, e.g., HI4PI; \citealt{HI4PI2016}) and CO \citep{dame2001} trace cold dust.
Another key observable is the reddening or extinction measurements of starlight, which occurs when their light passes through dust regions \citep[e.g.,][]{lenz:2017,lallement:2022,edenhofer:2024}. Finally, as recently shown in a companion paper by \cite{CG02_05}, the \ion{C}{ii} emission line at 158\,$\mu$m, as for instance measured by \COBE-FIRAS \citep{mather:1994}, traces hot dust regions. Before \cite{CG02_05}, it was not known to us that \ion{C}{ii} could be used directly as a high-precision dust tracer; however, as it is expected to be present in warmer regions of the Milky Way, it is not unexpected that this acts as a tracer for hot dust.\footnote{It has also been hypothesized that dust may be responsible for the \ion{C}{ii} deficit in luminous-FIR galaxies \citep{2020CII,2017Herschel_FIR_CII}.} 

Dust grains absorb radiation at wavelengths close to their particle size and re-emit as an imperfect blackbody. Historically, this has been modeled with a modified blackbody spectrum (MBB) characterized by a temperature, $T$, and a `tilt' to the spectrum, called the spectral index and denoted by $\beta$. These, along with an amplitude ($a$), define how much radiation at a particular frequency, $\nu$, is produced \citep[see, e.g.,][for a recent review]{Hensley2023}. During the \planck\ analysis \citep[][]{planck2016-l01,planck2014-a12}, Galactic dust was often modelled with a single MBB fit for each pixel, or direction, in the sky. However, there is evidence to suggest that there are different dust populations \citep{finkbeiner1999}, and with enough data it should be possible to make distinct templates modelling each independently. A significant challenge that \planck\ faced in their analysis was that there was insufficient data to distinguish these populations; in particular, there is a clear degeneracy between the dust temperature and the dust spectral index maps \citep{juvela:2012,planck2014-a12}.

In the current paper, we revisit the model introduced by \citet{CG02_05}. In that paper, it was through linear regression found that thermal dust emission in the \Planck\ HFI \citep{planck2016-l03} and \COBE-DIRBE \citep{hauser1998} data on large angular scales could be well described by a sum of five external dust templates, namely 1) \COBE-FIRAS \ion{C}{ii} \citep{fixsen:1998}; 2) \ion{H}{i} \citep{HI4PI2016}; 3) \Gaia\ extinction \citep{edenhofer:2024}; 4)Wisconsin H-$\alpha$ Mapper (WHAM) \Ha\ \citep{wham:2003,2016WHAM}; and 5) CO\,$J$=$1-0$ \citep{dame2001}. It was furthermore noted that the \ion{C}{ii} map template was associated with a higher temperature than the \ion{H}{i}  and CO templates; that the nearby \Gaia\ extinction template SED peaked between the hot and cold SEDs; and that the \Ha\ template was observed in absorption (i.e., having a negative template amplitude).

The main goal of the current paper is to establish a corresponding  multi-component model for \Planck\ HFI at full angular resolution. Since the FIRAS \ion{C}{ii} map has a very low angular resolution of $7^{\circ}$ FWHM, this means that the morphology of the hotter dust component has to be derived directly from the \Planck\ data, rather than traced by an already existing template. Furthermore, the \ion{H}{i}  and CO templates also have issues that prevent their direct use as fixed template (e.g., lower resolution and/or incomplete sky coverage). The key operational step in the current analysis is therefore to establish separate maps of cold and hot dust directly from the \Planck\ HFI data themselves. Since we do not use a template for the hot and cold dust, this also acts as a confirmation of the effects seen in \cite{CG02_05}, as the \planck\ data set is independent of the DIRBE data set, both in terms of the frequencies observed and the telescope. Finding cold and hot dust components tightly tracing \ion{H}{i} and \ion{C}{ii} gas acts as a strong confirmation of the results previously shown.

To summarize, following \cite{CG02_05}, we adopt the map produced by \citet{edenhofer:2024} with \textit{Gaia} data \citep{gaia:2016} to trace nearby dust in the Milky Way, and the \Ha\ map produced by \citet{wham:2003,2016WHAM} to trace a sub-dominant level of dust absorption. These two data sets are used as templates, fixing the relative amplitudes between pixels, but allowing the overall amplitudes of the maps, as well as the temperature and spectral indices, to vary globally. We introduce two additional dust components, here called the hot and the cold dust, which have freely varying amplitudes per pixel, but a global freely fitted temperature and spectral index. Ideally, we would have preferred to fit two cold dust components, one tracing \ion{H}{i} and another tracing CO as in the analysis of \citet{CG02_05}; however, because the SEDs of these two components are very similar, the \Planck\ HFI data do not allow for a clean separation of these two populations. For this reason, we fit only one effective cold dust component, which then accounts for both \ion{H}{i} and CO correlated dust. Compared to the previous Bayesian parametric \planck\ analysis \citep{planck2014-a12}, which had three free parameters per pixel ($\vec{T}$, $\vec{\beta}$ and $\vec{a}$ at each pixel), our model has ten global parameters and only two free parameters per pixel, namely the hot and cold dust amplitudes. 

This paper is one of seven companion papers from the \cosmoglobe\footnote{\url{https://cosmoglobe.uio.no}} data release 2 \citep[DR2;][]{CG02_01} reanalyzing the \COBE-DIRBE data \citep{hauser1998}. Of particular interest for this release is new constraints on the spectrum of the CIB \citep{CG02_03}, and better constraints on all foregrounds ranging from the microwave to the infrared, including zodiacal light \citep{CG02_02} and stars \citep{CG02_04}. Finally, in \cite{CG02_07} the new four-component dust model introduced in this paper is applied to the \COBE-DIRBE data.

The rest of the paper is organized as follows. In \cref{sec:data}, we discuss the data sets we use for this analysis. We present the sky and data model in \cref{sec:skymodel}. In \cref{sec:method}, we review the methodology used in this paper, primarily the Gibbs sampling and grid search tests, along with the optimization strategy used for the grid search. In \cref{sec:results}, we present the best-fit results for our new thermal dust model. In \cref{sec:gof}, we discuss the goodness of fit of the model, and, in \cref{sec:correlations}, the correlations with external line emission maps. In \cref{sec:mbb}, we compare to the previous \planck\ results and show a corresponding one-component MBB fit to our new multi-component dust model. Finally, in \cref{sec:conc}, we present our conclusions. In \cref{app:RMSScale,app:dustTemplates,app:residuals,app:GridSearchTests,app:dustCharacterization} we include supplementary details regarding the data preprocessing, dust templates, residuals, the grid search, and the dust characterization for clarity.

\begin{figure}
	\centering
	\includegraphics[width=0.9\linewidth]{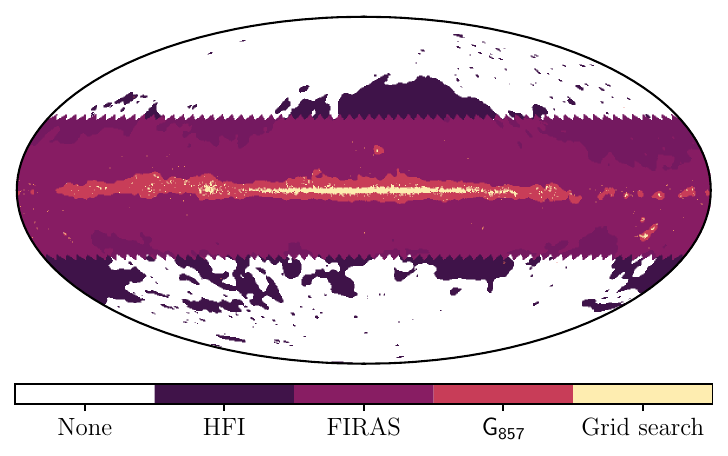}
	\caption{All masks used for different stages of the analysis. \emph{HFI:} Mask used during the HFI Gibbs sampling routine for the monopoles. \emph{FIRAS:} Mask used during the FIRAS Gibbs sampling routine for the monopoles. \emph{$\mathsf{G_{857}}$:} Mask used during the 857~GHz gain ($\mathsf{G_{857}}$) Gibbs sampling routine. \emph{Grid search:} Mask used to compare the $\chi^2$ in the grid search and when calculating correlations and summary statistics.}
	\label{fig:all_masks}
\end{figure}

\section{Data}
\label{sec:data}
Motivated by the work of \cite{CG02_05}, the following analysis uses the \planck\ PR4 data (\cref{sec:planckhfi}), several FIRAS maps (\cref{sec:firas}), and additional ancillary data (\cref{sec:otherdata}) to constrain a new high resolution four component dust model. We discuss our processing masks in \cref{sec:masks}. 

\subsection{\Planck\ HFI}
\label{sec:planckhfi}
The \planck\ PR4 (also known as \npipe) maps are the final \planck\ data release, improving on previous data releases by jointly analyzing the low frequency instrument (LFI) and the  high frequency instrument (HFI) data in a single pipeline, in addition to numerous other improvements (for details, see \citealt{planck2020-LVII}). 
For this analysis we use the single-horn HFI maps, since these maximizes our ability to separate thermal dust from CO line emission, while at the same time minimizes our sensitivity to polarization leakage. Specifically we use the 100--\{1-4\}, 143--\{1-7\}, 217--\{1-8\}, 353--\{1-8\}, 545--\{1,2,4\}, and 857--\{1-4\} horn maps,\footnote{The 143-8 and 545-3 maps were never released by the \planck\ collaboration due to non-Gaussian noise in the data, known as random telegraphic signal \citep{planck2013-p03}.} where the 100\,GHz and 143\,GHz maps are pixelized with a HEALPix \citep{healpix,Zonca2019} resolution of $N_{\mathrm{side}}=2048$, and higher frequency maps are pixelized with $N_{\mathrm{side}}=4096$.

In a pre-processing step we correct all maps for the stationary zodiacal light emission (which, unlike in earlier \Planck\ releases, is not removed in the PR4 processing), using the \Planck\ PR2 zodiacal light model \citep{planck2014-a09, planck2013-pip88, maris2006c}, and apply a correction for the CIB to the 857, 545 and 353\,GHz maps from the GNILC CIB results \cite{planck2016-XLVIII}. During the course of this work, the published single-horn frequency niose RMS maps were discovered to have residual offsets from the analysis pipeline, evident when comparing the map and RMS power-spectra, and we therefore apply a scaling correction to the instrumental noise root-mean squared (RMS) single-horn frequency maps, such that the tail of the maps' power spectra match the white-noise level from the RMS maps. These scaling values can be found in \cref{app:RMSScale}. We adjust the single-horn gain values for the 545\,GHz maps during a pre-processing step, with the values in \cref{tab:gains}, otherwise the gains are not adjusted, aside from the 857\,GHz gains, which are fit during the analysis (see \cref{sec:sampling}).

\subsection{FIRAS}
\label{sec:firas}
FIRAS was launched in 1989 as part of the \COBE\ satellite mission, with the goal of measuring the blackbody spectrum of the CMB to unprecedented precision \citep{fixsen1997}. The FIRAS maps have very low calibration uncertainties, and so we use them to help calibrate the map amplitudes. The FIRAS 857, 1251, 1809, 2081, 2135 and 2802\,GHz maps are used for this analysis in conjunction with the HFI maps. 

\subsection{Ancillary data}
\label{sec:otherdata}
The spatial distribution of the nearby dust amplitude is fixed to the \GAIA-based dust extinction template produced by \citet{edenhofer:2024}, which covers distances up to 1.25\,kpc. This primarily uses \GAIA\ distances and extinction estimates to build a 3D map of the nearby dust from \citet{2023Zhang} and \GAIA\ \citep{2019GAIA,2023gaia}, which also leverages the two Micron All Sky Survey (2MASS; \cite{2006AJ_2mass}), the Wide-field Infrared Survey Explorer (WISE and unWISE; \cite{allwise_ES,2010AJ_wise}), and the Large sky Area Multi-Object fibre Spectroscopic Telescope (LAMOST; \cite{2023yCatp038060301X_lamost,2022A&A_lamost,2022ApJS_lamost,2022yCat_lamost}). We use the total integrated dust map as our model in this analysis. 
Additionally, we use the Wisconsin H-$\alpha$ Mapper data \citep{wham:2003,2016WHAM} as a template for the \Ha\ correlated dust. Both maps are shown in \cref{app:dustTemplates}. The spectral parameters derived by \citep{CG02_05} from the combination of HFI and DIRBE data were used as fiducial values for the \Ha\ MBB, since we find that these parameters are poorly constrained by the \planck\ bands alone (see \cref{sec:results}). As for DIRBE, the \Ha\ dust appears with a negative sign in the current analysis, and therefore represents a small extinction component also in the \Planck\ bands.  
When considering our summary statistics in \cref{sec:correlations} we compare the new dust templates with the 157.7\,$\mu$m \ion{C}{ii} line emission as measured by the FIRAS, and the HI4PI data \citep{HI4PI2016}. 

\subsection{Masks}
\label{sec:masks}
When computing the $\chi^2$ for the grid search (see \cref{sec:statistics}) we apply a mask to remove the extreme outliers in the $\chi^2$ map, as can be seen in yellow in \cref{fig:all_masks}. This has $f_\mathrm{sky}=0.99$, removing only 1\% of the sky.
During the Gibbs sampling (see \cref{sec:sampling}), we apply a $|b|<30^\circ$ Galactic plane mask to the FIRAS maps when computing monopole estimates (representing an $f_\mathrm{sky}$ of 50\%), and a large Galactic cut (of areas with the largest foregrounds) to the HFI maps with an $f_\mathrm{sky}$ of 48\%, as can be seen in dark and light purple in \cref{fig:all_masks}. When estimating the gain for the 857\,GHz maps we employ a small mask of $f_\mathrm{sky}$ of 95\%, as seen in orange in \cref{fig:all_masks}. Any other component estimation (as discussed in \cref{sec:skymodel}) was done using the full sky.

\section{Data model}
\label{sec:skymodel}
The data model for this analysis reads
\begin{align}
    \vec{d}=&\mathsf{G}\mathsf{B}\sum_c \mathsf{M}_{c}\vec{a}_{c}+\vec{n}\nonumber\\
    \equiv & \vec{s}+ \vec{n},
\label{equ:dataModel}
\end{align}
where $\mathsf{G}$ is the instrumental gain matrix for each detector, $\mathsf{B}$ is the instrumental beam convolution, $\mathsf{M}_{c}$ is the foreground mixing matrix, which extrapolates the sky component ($c$) to the given frequency (including the bandpass and the spectral shape of the particular component) multiplied by amplitudes $\vec{a}_{c}$ per pixel, and $\vec{n}$ is the noise. As a reminder, we have pre-subtracted the zodiacal light and CIB from $\vec{d}$. Unlike in the other papers, however, we have not subtracted the CMB component, and so that is included in our sky model.

The \planck\ HFI bands contain several relevant sky components in addition to dust, such as free-free, CO, and the CMB.  
 As such, the sky model used for this analysis may be written as a vector per pixel
\begin{align}
    \vec{s}_\nu=&u_\nu 
    \left[\vphantom{\left(\frac{\nu}{\nu_\mathrm{d}}\right)^{\beta_\mathrm{d}+1}}(\vec{a} _\mathrm{CMB} + \vec{a}_{\mathrm{quad}}(\nu))\gamma(\nu) \right. &\mathrm{(CMB)} \nonumber \\
    &+\vec{t}_\mathrm{ff}\left(\frac{\nu_{0,\mathrm{ff}}}{\nu}\right)^2\frac{g_\mathrm{ff}(\nu;T_\mathrm{e})}{g_\mathrm{ff}(\nu_{0,\mathrm{ff}};T_\mathrm{e})} & \textrm{(Free-free)}\nonumber \\
    &+\vec{a}_\mathrm{cold}\left(\frac{\nu}{\nu_\mathrm{cold}}\right)^{\beta_\mathrm{cold}+1}\left(\frac{e^{h\nu_\mathrm{cold}/k_\mathrm{B}T_\mathrm{cold}}-1}{e^{h\nu/k_\mathrm{B}T_\mathrm{cold}}-1}\right) &\textrm{(Cold\ dust)}\nonumber \\
    &+\vec{a}_\mathrm{hot}\left(\frac{\nu}{\nu_\mathrm{hot}}\right)^{\beta_\mathrm{hot}+1}\left(\frac{e^{h\nu_\mathrm{hot}/k_\mathrm{B}T_\mathrm{hot}}-1}{e^{h\nu/k_\mathrm{B}T_\mathrm{hot}}-1}\right) &\textrm{(Hot\ dust)}\nonumber \\
    &+a_\mathrm{near}\vec{t}_\mathrm{near}\left(\frac{\nu}{\nu_\mathrm{near}}\right)^{\beta_\mathrm{near}+1}\left(\frac{e^{h\nu_\mathrm{near}/k_\mathrm{B}T_\mathrm{near}}-1}{e^{h\nu/k_\mathrm{B}T_\mathrm{near}}-1}\right)&\textrm{(Nearby\ dust)}\nonumber \\
    &+a_\mathrm{H\alpha}\vec{t}_\mathrm{H\alpha}\left(\frac{\nu}{\nu_\mathrm{H\alpha}}\right)^{\beta_\mathrm{H\alpha}+1}\left(\frac{e^{h\nu_\mathrm{H\alpha}/k_\mathrm{B}T_\mathrm{H\alpha}}-1}{e^{h\nu/k_\mathrm{B}T_\mathrm{H\alpha}}-1}\right) &\textrm{(H$\alpha$\ dust)}\nonumber \\
&+\vec{a}^{100}_\mathrm{co}h_\nu^{100}+\vec{a}^{217}_\mathrm{co}h_\nu^{217} +\vec{a}^{353}_\mathrm{co}h_\nu^{353}\left. \vphantom{\left(\frac{\nu}{\nu_\mathrm{d}}\right)^{\beta_\mathrm{d}+1}}\right] & \textrm{(CO)}\nonumber\\
    &+m_\nu, &\textrm{(Monopole)}
\label{equ:model}
\end{align}
where $u_\nu$ is a unit scaling conversion factor to go from brightness temperature to thermodynamic temperature, $\vec{a}_{\mathrm{quad}}(\nu)$ represents a relativistic quadrupole correction \citep{Notari:2015}, $\gamma(\nu)$ represents the CMB fluctuation spectrum in brightness temperature (including the CMB dipole and associated effects),
$\vec{a}$ are the amplitudes sampled per-pixel for the CMB ($\vec{a}_\mathrm{CMB}$), cold dust ($\vec{a}_\mathrm{cold}$), hot dust ($\vec{a}_\mathrm{hot}$), and the CO lines ($\vec{a}^{\{100,217,353\}}_\mathrm{co}$). We keep the CO line-ratios ($h_\nu^{\{100,217,353\}}$) fixed to the \planck\ PR4 values \citep{planck2020-LVI} for this analysis. We also keep the free-free component fixed due to the low constraining power of the HFI bands on the free-free emission. The template we use has an electron temperature of $T_e= 7000\,$K and amplitudes ($\vec{t}_\mathrm{ff}$) and Gaunt factor ($g_\mathrm{ff}$) from the \BeyondPlanck\ analysis \citep{bp01}. The amplitude scale values, $a_\mathrm{near}$ and $a_\mathrm{H\alpha}$ are multiplied by the templates ($\vec{t}_\mathrm{near}$ based on \citet{edenhofer:2024} and $\vec{t}_\mathrm{H\alpha}$ based on \citet{wham:2003,2016WHAM}) for the nearby dust ($\mathrm{near}$) and \Ha\ correlated (\Ha) dust respectively.   
Finally, the monopoles~($m_\nu$) are sampled for each frequency band. We omit synchrotron and anomalous microwave emission (AME), which are subdominant in the HFI frequency range. 
Note that, while in \cite{CG02_05} we fit a five-component dust model, of which one is the carbon monoxide correlated dust, in this case, we only have the four-components. Since the true CO emission is by definition spatially highly correlated with the CO-correlated dust component and since we do not include the DIRBE channels, the CO-correlated dust component is effectively merged with the cold dust component. 

\begin{figure*}[htbp]
    \centering
    \includegraphics[width=\linewidth]{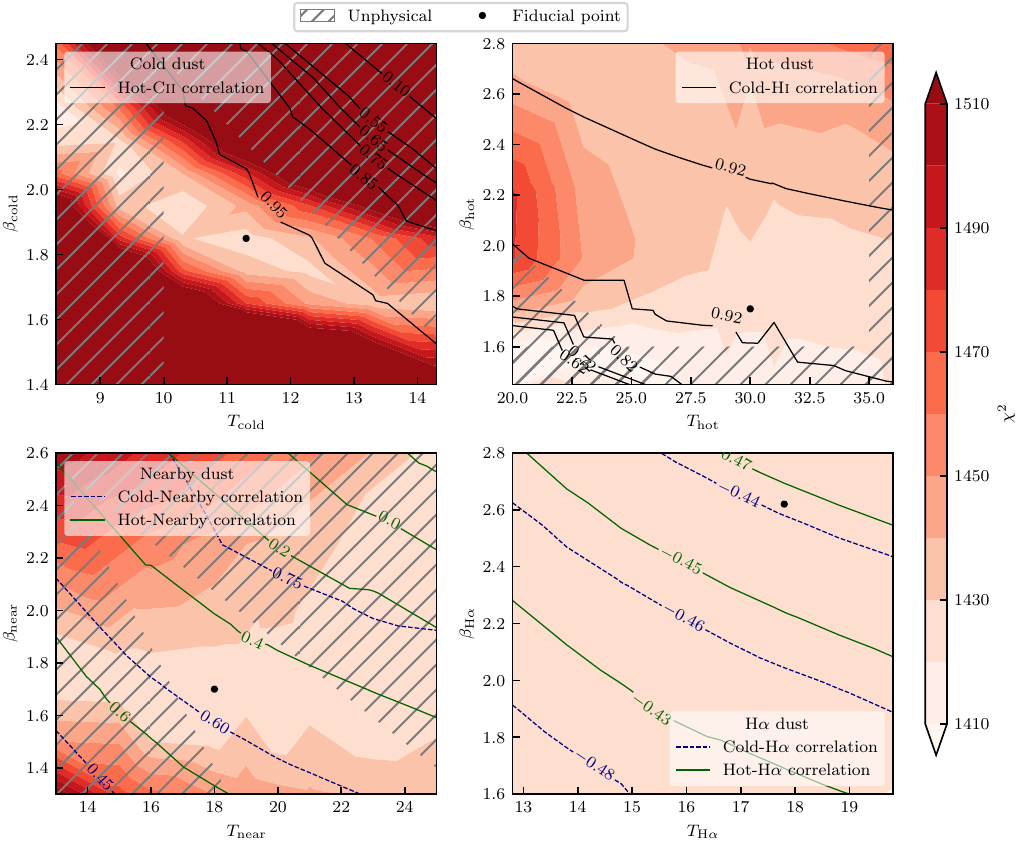}
    \caption{Contour plot of relevant statistical tests and $\chi^2$ (in red) for (clockwise from top left) cold dust, hot dust, \Ha\ dust and nearby dust. The black line shows the fiducial point used for all parameters and grey grids mark the unphysical regions of the parameter space. Maps of a subset of the grid points can be found in \cref{app:GridSearchTests}. }
    \label{fig:SEDGridContours}
\end{figure*}

\begin{figure}[htbp]
    \centering
    \includegraphics[width=\linewidth]{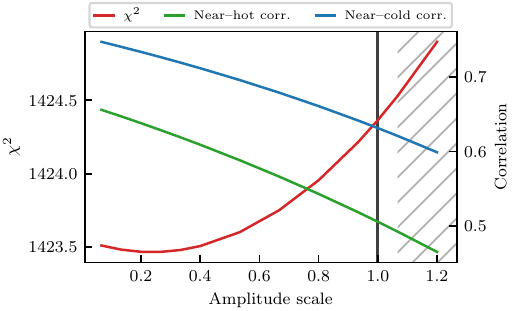}\\
    \includegraphics[width=\linewidth]{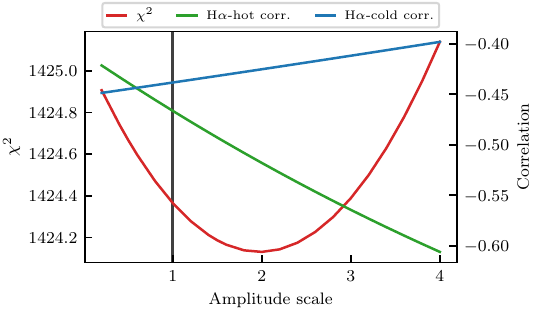}
    \caption{$\chi^2$ (in red) as a function of nearby (\emph{top}) and \Ha\ correlated (\emph{bottom}) dust amplitude scale. Green and blue curves show correlations with key companion components. Note that the range of the scale of the $\chi^2$ axes are very small. }
    \label{fig:SEDGridAmplitude}
\end{figure}

\begin{figure*}[t]
    \centering
\includegraphics[width=0.7\linewidth]{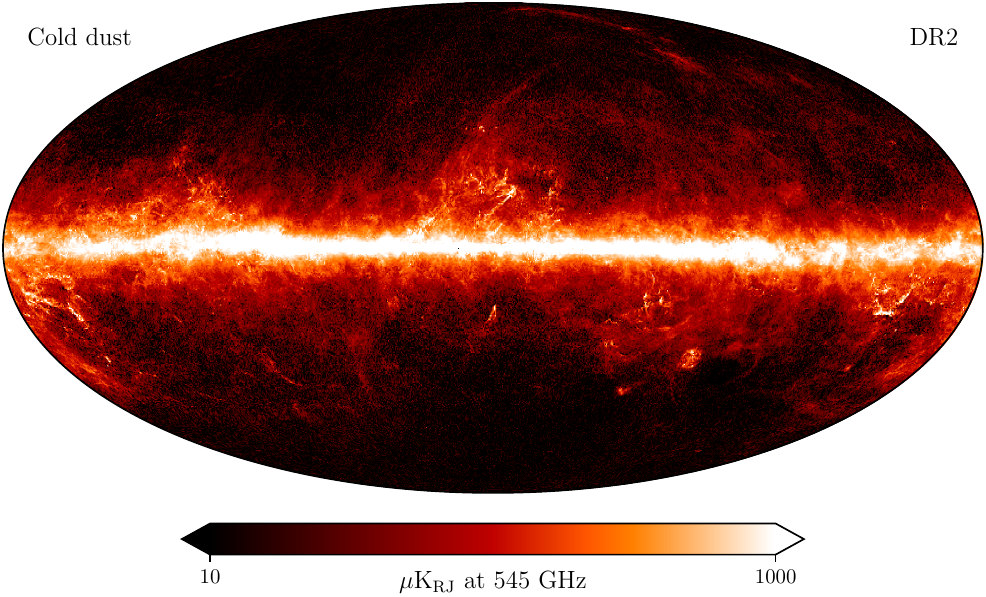}
\includegraphics[width=0.7\linewidth]{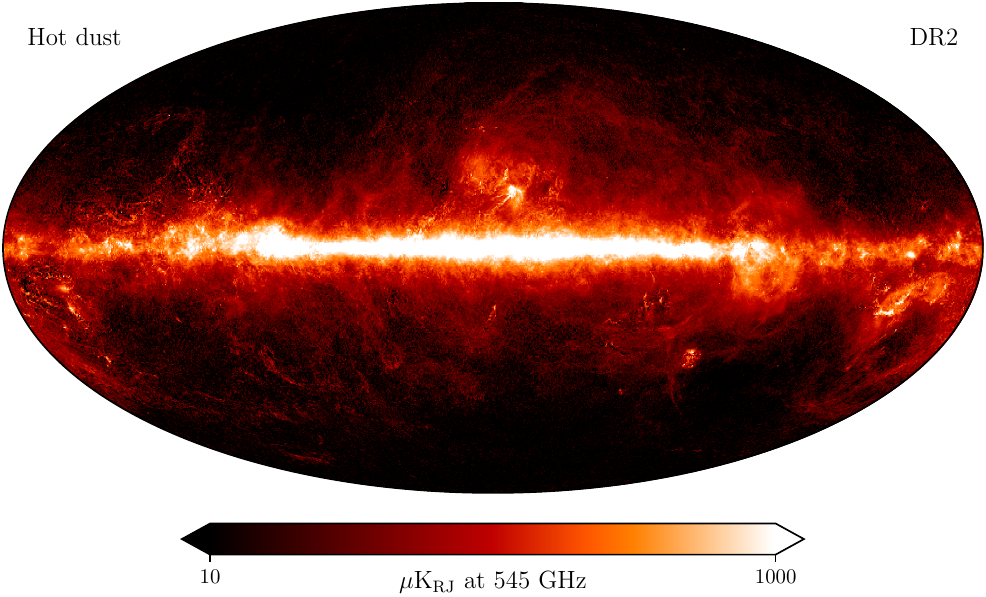}
\includegraphics[width=0.4\linewidth]{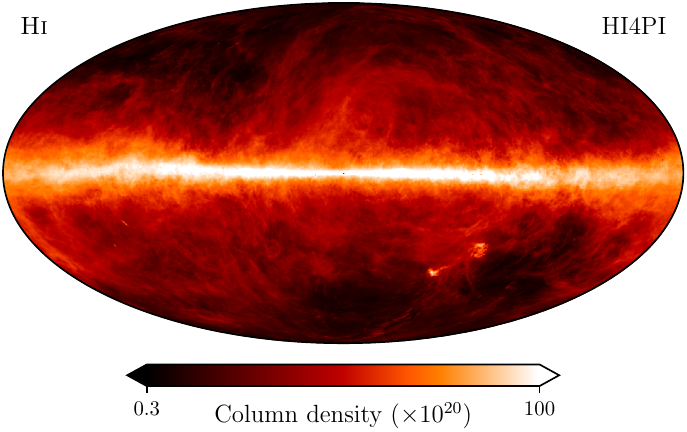}
\includegraphics[width=0.4\linewidth]{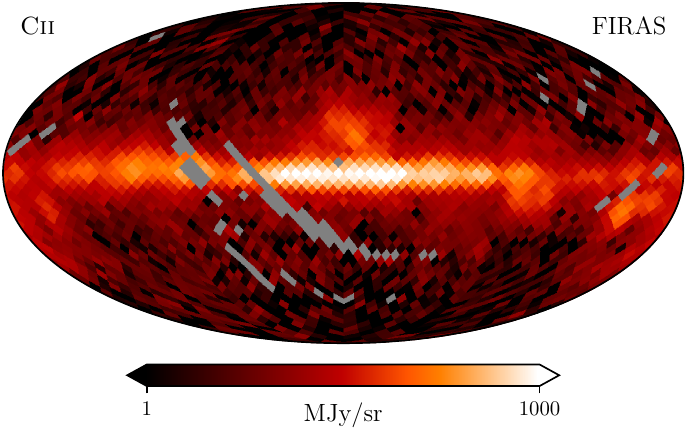}
    \caption{DR2 Cold and hot dust components (top panels) as compared with relevant tracers \ion{H}{i} and \ion{C}{ii} (bottom panels).}
    \label{fig:dust_figures}
\end{figure*}

\begin{figure}[htbp]
    \centering
    \includegraphics[width=0.9\linewidth]{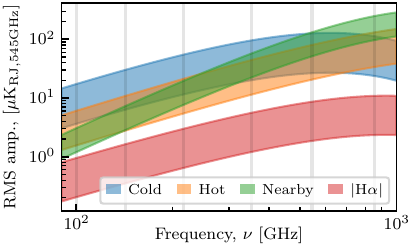}
    \caption{Spectral energy density (SED) for the four dust components across the HFI bands. Note that the \Ha\ dust is a negative amplitude since it is an extinction component, but the absolute value is plotted for clarity. The lower bound is set by a 70\% $f_\mathrm{sky}$ and the upper bound by a 94\% $f_\mathrm{sky}$. The grey vertical lines are the locations of the six \planck\ HFI bands. }
    \label{fig:dustSED}
\end{figure}

\begin{table}[tbp!]
\caption{Spectral parameters for the fiducial model found in the grid search.}
\begingroup
\newdimen\tblskip \tblskip=5pt
\nointerlineskip
\vskip -5mm
\footnotesize
\setbox\tablebox=\vbox{
     \newdimen\digitwidth
 \setbox0=\hbox{\rm 0}
 \digitwidth=\wd0
 \catcode`*=\active
 \def*{\kern\digitwidth}
  \newdimen\dpwidth
  \setbox0=\hbox{.}
  \dpwidth=\wd0
  \catcode`!=\active
  \def!{\kern\dpwidth}

\halign{\tabskip 0pt \hbox to 0.4750\linewidth{#\leaderfil}\tabskip 0pt&\hbox to 0.175\linewidth{\hfil#\hfil}\tabskip 0pt&\hbox to 
0.175\linewidth{\hfil#\hfil}\tabskip 0pt\cr
\noalign{\doubleline\vskip 1pt}
\omit\hbox to 1in{Component\hfil}& $T$ (K) & $\beta$\cr
\noalign{\vskip 4pt\hrule\vskip 6pt}
\noalign{\vskip 3pt}
Cold dust& $ 11\pm1 $ & $ 1.85 \pm 0.1$\cr
\noalign{\vskip 3pt}
Hot dust& $ 30 \pm 3$ & $ \ge 1.75 $\cr
\noalign{\vskip 3pt}
\Ha\ dust& $ 17.8^\mathrm{a} $ & $ 2.62^\mathrm{a} $\cr
\noalign{\vskip 3pt}
Nearby dust& $ 18 \pm 2$ & $ 1.7 \pm 0.1 $\cr
\noalign{\vskip 3pt}
\noalign{\vskip 3pt\hrule\vskip 3pt}}}
\endPlancktable
\endgroup
\tablewidth=\columnwidth
\tablenote {{\rm a}} { \footnotesize{\Ha\ dust is poorly constrained by the HFI data, and we therefore fix these parameters to the best-fit HFI+DIRBE result of \citet{CG02_05}.}}\par
\vskip -5pt
\label{tab:SEDs}
\end{table}

\begin{table}[tbp!]
\caption{Gain values for 545 and 857\,GHz bands.}
\begingroup
\newdimen\tblskip \tblskip=5pt
\nointerlineskip
\vskip -6mm
\footnotesize
\setbox\tablebox=\vbox{
\halign{\tabskip 0pt \hbox to 0.35\linewidth{#\leaderfil}\tabskip 0pt&
\hbox to 0.15\linewidth{\hfil#}\tabskip 0pt&
\hbox to 0.15\linewidth{\hfil#}\tabskip 0pt&
\hbox to 0.15\linewidth{\hfil#}\tabskip 0pt&
\hbox to 0.15\linewidth{\hfil#}\tabskip 0pt\cr
\noalign{\doubleline\vskip 1pt}
Band& 545-1 & 545-2 & & 545-4\cr
Gain scale$^a$& 0.99145 & 1.00487 & & 1.00830\cr
\noalign{\vskip 4pt\hrule\vskip 6pt}
\noalign{\vskip 3pt}
Band & 857-1 & 857-2 & 857-3 & 857-4\cr
Gain scale$^b$ & 0.99362 & 1.10242 & 1.02393 & 1.13470\cr
\noalign{\vskip 3pt}
\noalign{\vskip 3pt\hrule\vskip 4pt}}}
\endPlancktable
\endgroup
\tablewidth=\columnwidth
\tablenote {{\rm a}} { \footnotesize{The 545\,GHz gains from \npipe\ have a total uncertainty of 0.6\,\% \citep{planck2020-LVII}.}}\par
\tablenote {{\rm b}} { \footnotesize{The 857\,GHz gains sampled from the grid search and in the Gibbs sampling show uncertainties at the 0.001 level; note that this is conditional on the assumed gains at other frequencies.}}\par
\vskip -10pt
\label{tab:gains}
\end{table}

\begin{figure}[tbp]
    \centering
    \includegraphics[width=0.85\linewidth]{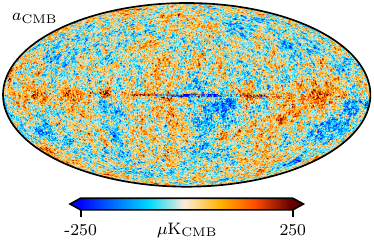}
    \includegraphics[width=0.85\linewidth]{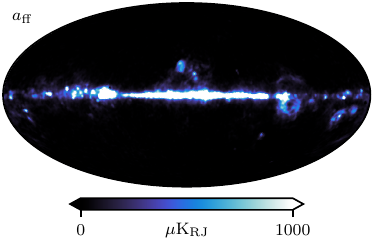}
    \includegraphics[width=0.85\linewidth]{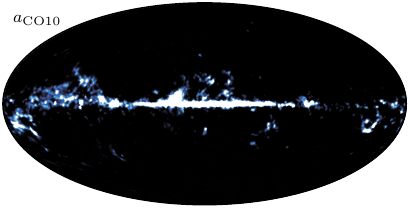}
    \includegraphics[width=0.85\linewidth]{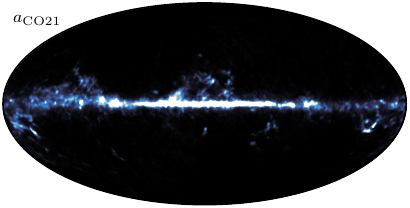}
    \includegraphics[width=0.85\linewidth]{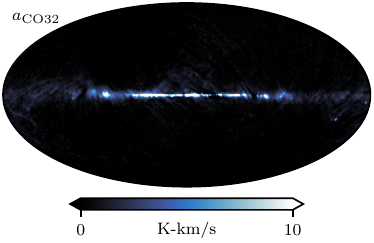}
    \caption{Sky model amplitude maps corresponding to the baseline thermal dust model. From top to bottom, the panels show CMB, free-free, CO $J=1-0$, CO $J=2-1$, and CO $J=3-2$. }
    \label{fig:SkyModel}
\end{figure}

\begin{figure*}[t]
    \centering
    \includegraphics[width=0.85\linewidth]{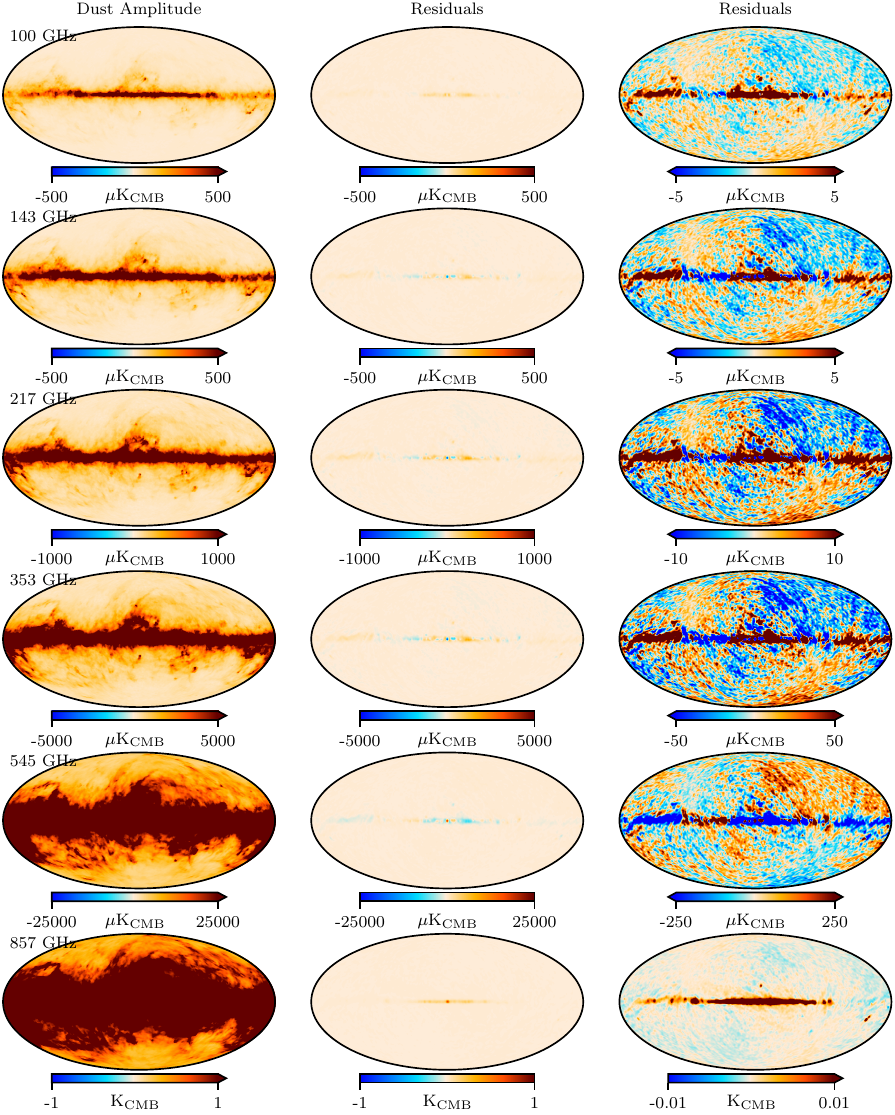}
    \caption{Left column shows HFI frequency maps from which all non-dust components have been subtracted, while the middle column shows the full residuals at each frequency using the same colour bar range. The right column shows the same residuals but with a 100 times smaller range. All maps have been smoothed to $2^{\circ}$ for clarity.}
    \label{fig:MapsVsResiduals}
\end{figure*}

\begin{figure}[t]
    \centering
    \includegraphics[width=\linewidth]{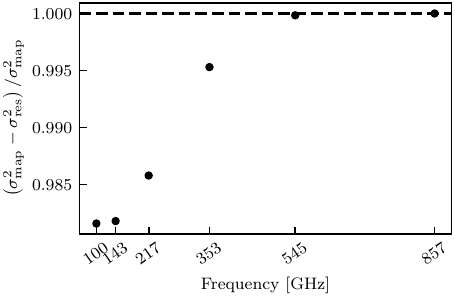}
    \caption{Dust model efficiency as a function of frequency as defined in terms of variance reduction.
	}
    \label{fig:efficiency}
\end{figure}

\begin{figure}[t]
    \centering
    \includegraphics[width=\linewidth]{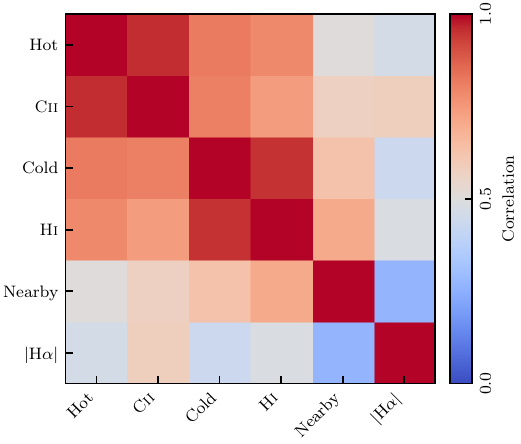}
    \caption{Pearson's correlation coefficient between dust components and tracers. Note how the correlation between the hot dust and the \ion{C}{ii} emission seen in FIRAS, as well as the cold dust and the HI4PI \ion{H}{i} emission, are very strong.
    }
    \label{fig:correlationSquare}
\end{figure}

\begin{figure}[t]
    \centering
    \includegraphics[width=0.97\linewidth]{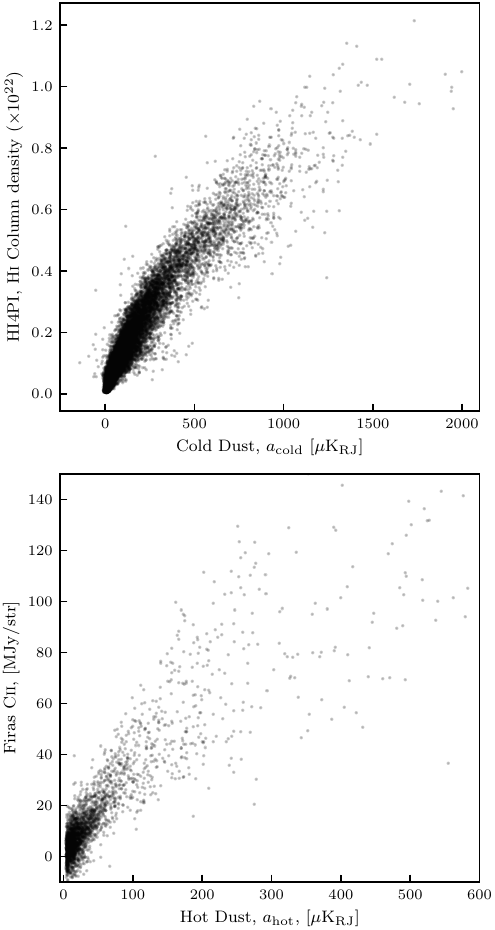}
    \caption{\emph{Top:} Scatter plot between the cold dust and HI4PI maps; both maps are smoothed to a common resolution of 16.2' FWHM, and pixelated at \nside=1024. \emph{Bottom:} Scatter plot between the hot dust and FIRAS \ion{C}{ii} maps; both maps are smoothed to a common resolution of 420' FWHM, and pixelated at \nside=16. 
    }
    \label{fig:hotCiiCorr}
\end{figure}

\begin{figure*}[t]
    \centering
    \includegraphics[width=0.8\linewidth]{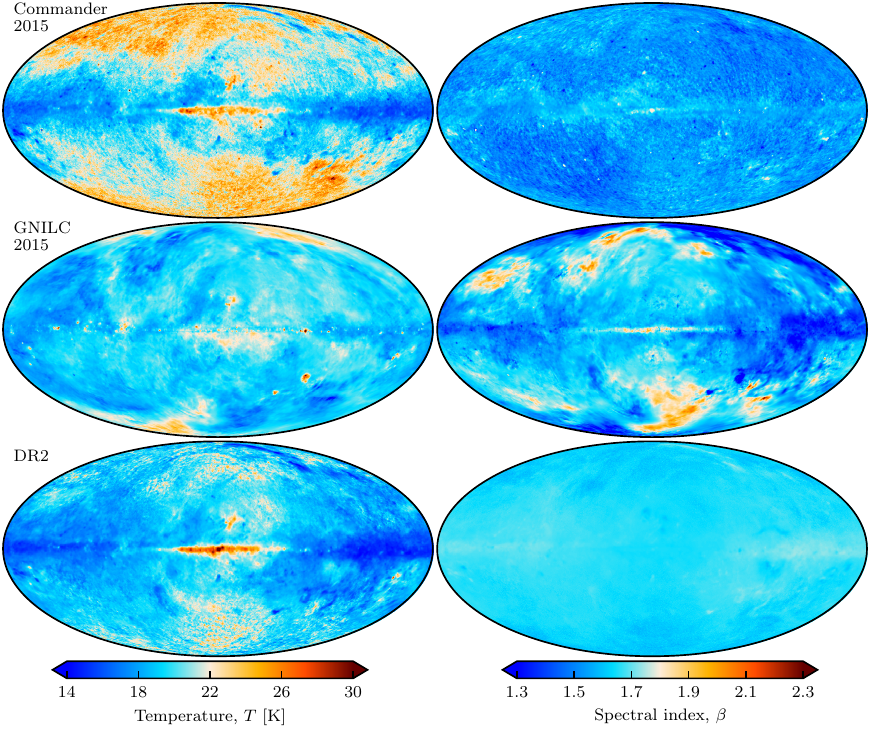}
    \caption{Comparison of the dust temperature $T$ and spectral index $\beta$ sky fits between \planck\ 2015, GNILC 2015 and the new \cosmoglobe\ DR2 results.}
    \label{fig:beta_T}
\end{figure*}

\section{Methods}
\label{sec:method}
We employ a two-step hybrid algorithm to constrain the above data model to the HFI data, namely a Bayesian Gibbs sampler to estimate pixel-by-pixel fit parameters, coupled to a parameter grid search for global parameters. The advantage of this approach is that it allows us to scan large parameter volumes with a much lower number of posterior evaluations than by Gibbs sampling alone, due to the strong correlations between the global and per-pixel parameters.
\subsection{Bayesian analysis and Gibbs sampling}
\label{sec:sampling}

For the per-pixel parameters in \cref{equ:dataModel,equ:model} we use a Bayesian Gibbs sampler called \texttt{Commander3} \citep{bp03}. Gibbs sampling \citep{geman:1984} is a Monte-Carlo sampling technique useful for very high dimensional parameter spaces, such as our per-pixel amplitudes, as it allows one to iteratively sample the various parameters in turn, instead of having to draw samples from the complete distribution. 
Our Gibbs sampler, for this case, employs the algorithm 
\begin{alignat}{2}
  \label{eq:gibbs1}
  \vec{a}_c &\leftarrow P(\vec{a}_s &&|{\bf d}, \phantom{\vec{a}_c ,} \mathsf{G_{857}},m)\\
\mathsf{G_{857}} &\leftarrow P(\mathsf{G_{857}} &&|{\bf d}, \vec{a}_c,\phantom{\mathsf{G_{857}},}m)\\ 
m &\leftarrow P(m &&|{\bf d}, \vec{a}_c,\mathsf{G_{857}}\phantom{,m}) ,
\label{eq:gibbs3}
\end{alignat}
where the arrow operator $\leftarrow$ indicates updating the variable on the left with a sample drawn from the conditional distribution on the right. First, the amplitudes for all the components $c$ (including dust, CMB, and CO, as discussed in \cref{sec:skymodel}) are sampled while holding the 857\,GHz gains~($\mathsf{G_{857}}$) and monopoles~($m$) fixed, and proceeding through the subsequent two steps before looping back to the first stage, obtaining samples for each free parameter. The amplitudes $\vec{a}_c$ are by far the most expensive step, as they represent per-pixel parameters across the full sky at high resolution, whereas the gains and monopoles are a single parameter per channel. For a detailed discussion of Gibbs sampling see \cite{bp01,eriksen2008} and references therein. This approach is the foundation of the \cosmoglobe\ framework, based on the \texttt{Commander} software \citep{eriksen:2004,seljebotn:2019,bp03}, used throughout this data release.

\subsection{Parameter grid search and optimization strategy}
\label{sec:statistics}
Having only access to HFI data, the four dust component amplitudes in the current model are highly correlated with their corresponding spectral parameters, $T$ and $\beta$. This results in a long Monte Carlo correlation length in a fully integrated Gibbs sampler, which then translates into a high computational cost for scanning large parameter volumes. To mitigate this, we instead perform an extensive grid search over the spectral parameter and the \Gaia\ and H$\alpha$ scaling amplitudes, in which we initialize a large number of independent Gibbs chains with different parameter combinations. For each fixed set of spectral parameters, we then run a short Gibbs sampler as given by Eqs.~\ref{eq:gibbs1}--\ref{eq:gibbs3} to fit the amplitudes per pixel of the hot and cold dust; the CMB and the CO amplitudes; the gains at 857\,GHz; as well as monopoles. We then compare the $\chi^2$ from the residuals, the correlations between the nearby-hot, nearby-cold, \Ha-hot and \Ha-cold maps, the correlation between the cold and HI4PI maps, and we inspect the hot and cold maps for large scale negative amplitudes. The nearby and H$\alpha$ dust components describe dust structures that are located physically close to the Earth, while the hot and cold components should account for everything not fit by those, i.e., more distant dust features. In general, a solution with a lower correlation between the nearby and H$\alpha$ components on the one hand, and hot and cold dust on the other hand, is therefore preferable over solutions with higher correlations. 

At the outset of this analysis, we do not know \emph{a-priori} the optimal grid range for each parameter for the currently available HFI dataset. As a first guess, we therefore start from the best-fit HFI+DIRBE parameters presented by \citet{CG02_05}, and start exploring a large but sparsely parameter volume around this position. We then identify the best-fit point of the resulting grid, based on the combination of $\chi^2$ and correlation statistics defined above, and restart the gridding around the new best-fit point. This process is repeated several times. Finally, we compute dense 1D or 2D grids around the final best-fit point for each global parameter. 

Both during the analysis and inspection phases, we impose two absolute physicality priors. First, we require that the large scale amplitudes of the hot and cold dust must both be positive, as expected for thermal dust emission. Second, we require all dust temperature $T$ to lie between 10 and 35\,K, noting that a wide range of previous studies strongly disfavour more extreme temperatures on large angular scales \citep{2017A&A_DustParam,2016SAA_DustReview,Hensley2021}. Parameter solutions which fail either of these criteria are shown as gray hatched regions in the following.

\section{Thermal dust template results}
\label{sec:results}

The global best-fit dust parameters derived through the above procedure are listed in \cref{tab:SEDs}.
The error bars in this table are derived from the final low-dimensional grids, although we note that these are only approximate; due to the extremely high signal-to-noise ratio of the \Planck\ HFI data, the final data-minus residuals are dominated by signal modelling errors rather than white noise, and a strict $\chi^2$-based uncertainty will vastly underestimate the true uncertainty. Instead, we define the effective uncertainty as the parameter range for which the $\Delta\chi^2 \sim 10$ with respect to the best-fit solution; we have confirmed visually that parameter combinations within this range correspond to both nearly indistinguishable and physically viable amplitude maps, as expected for a $1\,\sigma$ fluctuation.

Considering the specific numerical values listed in \cref{tab:SEDs}, we note that the hot dust component has a best-fit temperature of 30\,K, while the cold dust component has a temperature of 11\,K. These are both slightly more extreme than those reported for the corresponding components derived from DIRBE+HFI by \citet{CG02_05}, which were 26\,K for \ion{C}{ii}; 12\,K for CO; and 16\,K for \ion{H}{i}. We emphasize that the HFI frequency bands do not cover the peak of the SED for a $\gtrsim$12\,K MBB, and the derived HFI-only temperatures are therefore particularly susceptible to the 857\,GHz gain. As such, we do not attach a high statistical significance to the precise numerical values of the spectral parameters listed in \cref{tab:SEDs}, but rather note the main results in the current paper are not the MBB parameters as such (which are strongly degenerate with the 857\,GHz gain), but rather the morphology of the corresponding amplitude maps (which are not strongly degenerate with the 857\,GHz gain).

Next, \cref{fig:SEDGridContours} shows $(T,\beta)$ grids for each of the four dust components, computed by fixing all other parameters at the global best-fit solution. The underlying colors indicate the respective reduced $\chi^2$ distribution, while the contours show the Pearson correlation coefficient for a relevant companion map. Starting with the cold dust component in the top left panel, we first observe the well-known ``banana-shaped'' degeneracy between $T$ and $\beta$; a higher temperature can be compensated by a lower $\beta$, which together maintains a nearly unchanged SED peak position \citep[e.g.,][]{juvela:2012}. Here we clearly see that the HFI data alone do not constrain the cold dust SED well, as any temperature between 8 and $\sim$14\,K results in an acceptable $\chi^2$. However, visual inspection the resulting amplitude maps (as shown in Figs.~\ref{fig:hot_cold_dust_set}--\ref{fig:hot_hot_dust_ratio}) shows that higher temperatures lead to strong degeneracies with the hot dust component, and thereby a negative large-scale amplitude. The right-most hatched region shows values that are excluded due to these unphysical solutions, while the leftmost regions are excluded due to extreme temperatures. The contours in this plot shows correlations between FIRAS \ion{C}{ii} and the hot dust component, and here we see that any cold dust MBB parameter configuration in the preferred region result in a correlation coefficient higher than 0.95.

Moving on to the hot dust MBB parameters, shown in the top right panel, we first of all note that the $\chi^2$ distribution remains nearly constant over most of the plotted region. This is due to the fact that the hot dust component dominates at the 857\,GHz channel, and almost any change in the proposed SED can be compensated by an 857\,GHz gain adjustment combined with a small modification to the cold dust amplitude map. As a result, the morphology of the hot dust component is extremely stable with respect to both MBB and gain assumptions, up to an overall multiplicative re-scaling factor which is uncertain, and the resulting morphology correlates strongly with FIRAS \ion{C}{ii}, in full agreement with the observation made by \citet{CG02_05}.

The bottom left panel show a similar $(T,\beta)$ grid for the nearby dust component. In this case, the $\chi^2$ provides little constraining power in $T$, but non-trivial constraints in $\beta$. Furthermore, due to the extended nature of this component on the sky, only a relatively narrow wedge is allowed by the physicality priors; too high temperatures and spectral indices lead to negative hot and cold dust amplitudes in some parts of the sky, while too low temperatures and spectral indicies lead to negative amplitudes in other parts of the sky. A temperature around 18\,K and a spectral index around 1.7 constitute a sweet spot where all constraints are simultaneously fulfilled.

Finally, the bottom right panel shows the same quantities for the H$\alpha$ component. In this case, we see that both the $\chi^2$ and the correlations are virtually independent of the spectral parameters. This is due to a combination of the overall low amplitude of this component, and the fact that much of the template power is removed by the analysis mask. Given the current analysis setup, we therefore fix all H$\alpha$ parameters to the best-fit values derived in the DIRBE+HFI analysis by \citet{CG02_05}. At the same time, we also note that if we exclude this template entirely from the fit (as indeed was done in the early stages of the analysis), a negative residual appears with the same morphology as the WHAM template; its presence in the model does visually improve the overall fit, even though it is challenging to define a robust statistical sampling algorithm using HFI data alone.

\Cref{fig:SEDGridAmplitude} shows corresponding one-dimensional distributions for the overall scaling amplitudes, $a_i$, of the nearby and H$\alpha$ components. In both cases, we normalize the template scaling such that $a=1$ by construction, and then explore values around this to probe the overall stability. First, we note that the $\chi^2$ range is very small. Since the MBB temperature of both these components lie between those of the cold and hot dust components, nearly any change in $a_i$  can be countered by adjusting the cold and hot dust amplitude maps. At some point, however, too extreme variations lead to negative values in those freely fitted amplitude maps, and the resulting solution is excluded by the physicality prior. Specifically, values larger than $a_{\mathrm{near}} \gtrsim 1.05$ are excluded in the top panel of \cref{fig:SEDGridAmplitude}. We also note that values close to unity are preferred over, say, $a_{\mathrm{near}} \gtrsim 1.1$, due to the lower correlation with the hot and cold component map, implying a cleaner separation between the nearby, hot, and cold components.

The actual best-fit cold and hot amplitude maps are shown in the top two panels of \cref{fig:dust_figures}, while the two external templates are shown in \cref{fig:dust_template_figures}, and full resolution maps may be found on the \cosmoglobe\ products' webpage\footnote{\href{https://www.cosmoglobe.uio.no/products/cosmoglobe-dr2.html}{https://www.cosmoglobe.uio.no/products/cosmoglobe-dr2.html}}. The bottom two panels of \cref{fig:dust_figures} show the HI4PI and FIRAS \ion{C}{ii} maps. Already at a visual level, the correlations between these and the two amplitude maps in the top panels are striking: The cold dust component is strongly correlated with \ion{H}{i}, while the hot dust component is strongly correlated with \ion{C}{ii}, in agreement with \citet{CG02_05}. This will be discussed further in \cref{sec:correlations}. 

\Cref{fig:dustSED} shows the best-fit MBB fits as a function of frequency, with amplitudes set by the RMS amplitude of the map for sky cuts about the Galactic plane (94\% and 70\% $f_\mathrm{sky}$). This emphasizes both the challenge of doing such a fit with only HFI data (since within the bands fit by HFI the spectra are relatively flat) but also the incredible power additional data sets (such as the \cite{edenhofer:2024} and \cite{wham:2003,2016WHAM}) bring, allowing us to effectively separate these dust components. A different visualization of the SEDs (spectral energy densities), the relative fraction of each dust component at 545\,GHz, can be seen in \cref{fig:dustFrac} and discussion in \cref{app:dustCharacterization}.

Results for the gain estimation for the fiducial model can be seen in \cref{tab:gains}. The 857\,GHz gain values have formally a large impact on the overall dust model, since they shift the measured spectrum in those bands, and the higher frequencies give us the longest ``lever'' on the dust (see again \cref{fig:dustSED}). However, they are also associated with large intrinsic PR4 uncertainties at the 10\,\% level \citep{planck2020-LVII}, and this is a major contributor to our error budget on the dust parameters. 

The amplitude maps obtained for other components are shown in \cref{fig:SkyModel}. We recover the CMB with great accuracy even near the Galactic plane. The free-free is a fixed input template, and the recovery of the CO lines is sufficient for this analysis. The recovery of the CO line could be greatly improved in a future analysis where we allow the line ratios to fluctuate and  include \Planck\ LFI data to better control the low-frequency foregrounds. 

Finally, the monopole values are simply offsets for each map, which are effectively a correction term for any residual offset that might have been introduced during map-making, and as such are not presented here due to space constraints.

\section{Efficiency assessment and goodness of fit}
\label{sec:gof}
We assess the overall goodness of fit by comparing our model to the (data-minus-model) residuals in \cref{fig:MapsVsResiduals}. The left column shows frequency maps from which all non-dust components have been subtracted. The middle column shows the full data-minus-signal residuals at the same scale range as the maps, and the right column shows the residuals with a factor of 100 smaller range. While there is still some structure in the residual maps, they are largely driven by structures that appear uncorrelated with Galactic dust. This suggests that future improvements may be made to the gain and zodiacal light modelling; however, such corrections need to be made at the level of time-ordered data, and fall outside the scope of the current paper. Residuals for each individual detector can be found in \cref{app:residuals}.

To quantify the efficiency of the dust model, \cref{fig:efficiency} shows the ratio between the dust residual and full residual variance, normalized by the dust residual variance. We smooth the maps by a beam of 10$^\prime$ here but keep them at their maximum resolution per-band. If our dust model happened to be a perfect tracer of the true dust signal, the residual would be zero, and the efficiency ratio would be unity; if the dust model happened to not trace anything of the dust, the residual would be equal to the data, and the ratio would be zero. This statistic therefore quantifies the amount of signal variance explained by the model.

Overall, we see that the efficiency is very high, with more than 99\,\% of the 545 and 857\,GHz variance explained by the model. Even for the 100\,GHz channel, which is strongly affected by intrinsic CO emission, the ratio is still higher than 98\,\%. Overall, this model is an extremely efficient tracer of thermal dust across the full HFI frequency band, despite the fact that it has only two degrees-of-freedom per pixel.  

\section{Correlations with external line emission maps}
\label{sec:correlations}

The key result emerging in the low-resolution analysis of \citet{CG02_05} was that a large fraction of the large-scale thermal dust emission in HFI could be traced by the HI4PI \ion{H}{i} and FIRAS \ion{C}{ii} maps. In this paper, we have turned this observation around, and simply fitted two components per pixel with free global parameters, while imposing no priors on the resulting spatial distribution. Yet, in \cref{fig:dust_figures} we noted that the resulting sky maps indeed appeared visually similar to HI4PI and FIRAS. We now quantify these obsevations in terms of Pearson's correlation coefficient, as shown in \cref{fig:correlationSquare} for the full set of dust components and tracers.

First, we note that all maps are non-trivially correlated with all others. This is expected, given that they all trace the same overall Galactic ISM. Still, some combinations exhibit stronger correlations than other. Specifically, the hot and \ion{C}{ii} components correlate at the 95\% level, while the cold dust and HI4PI maps correlate at the 92\% level. For comparison, the hot and cold components correlate internally at the 80\% level. Correlations between \ion{C}{ii} and \ion{H}{i} have been known for a long time, and discussed regarding gas populations in the Milky Way Galaxy \citep{2014_CII,2010_CII,2013_CII}, so seeing the same effect between the hot and cold dust here is not unexpected.

\Cref{fig:hotCiiCorr} shows per-pixel scatter plots between our cold and hot dust maps and the respective primary physical tracers. The top panel shows the cold dust and HI4PI column density maps at $N_{\mathrm{side}}=1024$, and the bottom panel shows the hot dust and FIRAS \ion{C}{ii}, downgraded to the $7^\circ$ resolution of the FIRAS maps. 

\section{Modified blackbody fit to multi-component dust model}
\label{sec:mbb}

So far we have fit a four-component dust model to the \Planck\ HFI
data, where each component is modelled in terms of an MBB spectrum
with global\footnote{Global here meaning to a single value per dust map for the whole sky.} spectral parameters $T$ and $\beta$. In this section,
we compare this to the \Planck\ 2015 \citep{planck2014-a12} analysis,
where only one MBB dust component was fit to a combination of \Planck,
\WMAP\ \citep{bennett:2013}, and Haslam \citep{haslam1982} data, while
allowing spatial variations in $T$ and $\beta$. The \Planck\ 2015 maps
are plotted in the top panel of \cref{fig:beta_T}. In this case, both
$T$, $\beta$, and the dust amplitude are fitted per HEALPix
$N_{\mathrm{side}}=256$ pixel using \texttt{Commander1}
\citep{eriksen2008}.\footnote{\texttt{Commander1} is distinct from
\texttt{Commander3} in that it assumes a uniform beam across all
channels, while \texttt{Commander3} allows for multi-resolution data}
While $\beta$ is relatively flat on the sky, we see that $T$ varies
significantly. Note that in \citet{planck2014-a12}, the CIB was not
fitted separately, and the CIB fluctuations, therefore, mostly went
into the thermal dust component. The hot areas at high Galactic
latitudes correspond to CIB mixing in the dust component.

\citet{planck2016-XLVIII} fit separate CIB and thermal dust
components to the \Planck\ data using GNILC, a needlet internal linear combination code with
power spectrum priors. Here, the thermal dust component was also
modelled as a one-component MBB with variable $T$ and $\beta$. The corresponding $T$ and $\beta$ maps are plotted
in the middle row of \cref{fig:beta_T}. Compared to the \Commander\ maps
above, the CIB contamination is now gone from the $T$ map, but the $\beta$
map varies more on the sky.

In the bottom of \cref{fig:beta_T} we have used \Commander1 to fit a
one-component MBB with varying spectral indices to the four-component dust
model with spatially flat spectral parameters, as developed in this
paper. We see that an underlying simple four-component dust model appears
very similar to a one-component model with a nearly constant $\beta$
and a varying temperature $T$. Except at high latitudes, where the
\Planck\ 2015 $T$ map is CIB dominated and the current analysis
subtracts CIB fluctuations at the input map level, the two $T$ maps
agree very well. This shows that the effective one-component MBB from
\Planck\ 2015 can instead be replaced with the new, simpler, four-component
model, which in fact has fewer free parameters per pixel.

\section{Conclusions}
\label{sec:conc}

In \cite{CG02_05} we found that large-scale thermal dust emission in the microwave and infrared regimes may be accurately described by a multi-component model that includes \ion{H}{i} correlated dust, \ion{C}{ii} correlated dust, stellar extinction dust, CO correlated dust and \Ha\ correlated dust. Motivated by these results, in this paper we fit a corresponding four-component dust maps to the full-resolution \planck\ HFI data. This analysis results in two novel dust maps, corresponding to the hot and cold dust populations, respectively, which have global temperatures and spectral indices of $T_\mathrm{hot}=30$\,K, $\beta_{hot}=1.75$, and $T_\mathrm{cold}=11$\,K, $\beta_\mathrm{cold}=1.85$. The cold dust amplitude map has a 92\% correlation coefficient with the HI4PI map \citep{HI4PI2016}, while the hot dust component has a striking 95\% correlation with the FIRAS \ion{C}{ii} map. With only two free parameters per pixel, this dust model explains more than 99\,\% of the signal variance at 545 and 857\,GHz, and more than 98\,\% at 100\,GHz. Not only is this model statistically more compact than previous efforts, which ultimately should translate into stronger CMB and CIB fluctuation constraints from HFI, but the resulting component maps also trace known physical effects.  

When fitting a single MBB component to the sum of the four dust templates, we observe similar morphologies in the resulting spectral maps as in the \planck\ 2015 dust model, although without CIB contamination at high latitudes. This strongly suggests that previous analyses that modelled dust as a one-component MBB with spatially varying spectral indices have actually been averaging over different effects that ideally should not be averaged over. As such, by using multiple components with simpler spatial variations, our model is simpler, more physical, and more economical. Furthermore, it is computationally much easier to fit for linear (amplitude) parameters than for non-linear (spectral) parameters so this will also provide speed-ups to future dust simulations and analysis. 

The current model is based on isotropic SEDs. However, we expect there to be real variations in each component, simply because the dust is located at different distances to nearby stars and by simple geometry have varying temperatures. The current model is just a first large-scale approximation. On small scales there will be real variations, and in the future, this work should be extended through joint analyses with DIRBE, AKARI, WISE, SPHEREx, and others, to produce even higher resolution dust maps with greater sensitivity and longer frequency coverage. 
This result is also critically important for $B$-mode polarization experiments, as it may suggest that future $B$-mode polarization experiments should implement capabilities to disentangle hot and cold dust efficiently, which suggest requiring data with their highest frequency around 800-1000\,GHz.

\begin{acknowledgements}
 The current work has received funding from the European
  Union’s Horizon 2020 research and innovation programme under grant
  agreement numbers 344934 (YRT; CosmoglobeHD) and 351037 (FRIPRO; LiteBIRD-Norway), 819478 (ERC; \textsc{Cosmoglobe}) and 772253 (ERC;
  \textsc{bits2cosmology}). Some of the results in this paper have been derived using the HEALPix \citep{healpix} package.
  We acknowledge the use of the Legacy Archive for Microwave Background Data
  Analysis (LAMBDA), part of the High Energy Astrophysics Science Archive Center
  (HEASARC). HEASARC/LAMBDA is a service of the Astrophysics Science Division at
  the NASA Goddard Space Flight Center.  
  
   This publication makes use of data products from the Wide-field Infrared Survey Explorer, which is a joint project of the University of California, Los Angeles, and the Jet Propulsion Laboratory/California Institute of Technology, and NEOWISE, which is a project of the Jet Propulsion Laboratory/California Institute of Technology. WISE and NEOWISE are funded by the National Aeronautics and Space Administration.
   
   This work has made use of data from the European Space Agency (ESA) mission
{\it Gaia} (\url{https://www.cosmos.esa.int/gaia}), processed by the {\it Gaia}
Data Processing and Analysis Consortium (DPAC,
\url{https://www.cosmos.esa.int/web/gaia/dpac/consortium}). Funding for the DPAC
has been provided by national institutions, in particular the institutions
participating in the {\it Gaia} Multilateral Agreement.

This paper and related research have been conducted during and with the support of the Italian national inter-university PhD programme in Space Science and Technology. Work on this article was produced while attending the PhD program in PhD in Space Science and Technology at the University of Trento, Cycle XXXIX, with the support of a scholarship financed by the Ministerial Decree no. 118 of 2nd March 2023, based on the NRRP - funded by the European Union - NextGenerationEU - Mission 4 "Education and Research", Component 1 "Enhancement of the offer of educational services: from nurseries to universities” - Investment 4.1 “Extension of the number of research doctorates and innovative doctorates for public administration and cultural heritage” - CUP E66E23000110001.
\end{acknowledgements}

\bibliographystyle{aa}
\bibliography{../../common/CG_bibliography,references,Planck_bib}

\begin{thebibliography}{73}
\expandafter\ifx\csname natexlab\endcsname\relax\def\natexlab#1{#1}\fi

\bibitem[{{Ade} {et~al.}(2019){Ade}, {Aguirre}, {Ahmed}, {Aiola}, {Ali}, {Alonso}, {Alvarez}, {Arnold}, {Ashton}, {Austermann}, {Awan}, {Baccigalupi}, {Baildon}, {Barron}, {Battaglia}, {Battye}, {Baxter}, {Bazarko}, {Beall}, {Bean}, {Beck}, {Beckman}, {Beringue}, {Bianchini}, {Boada}, {Boettger}, {Bond}, {Borrill}, {Brown}, {Bruno}, {Bryan}, {Calabrese}, {Calafut}, {Calisse}, {Carron}, {Challinor}, {Chesmore}, {Chinone}, {Chluba}, {Cho}, {Choi}, {Coppi}, {Cothard}, {Coughlin}, {Crichton}, {Crowley}, {Crowley}, {Cukierman}, {D'Ewart}, {D{\"u}nner}, {de Haan}, {Devlin}, {Dicker}, {Didier}, {Dobbs}, {Dober}, {Duell}, {Duff}, {Duivenvoorden}, {Dunkley}, {Dusatko}, {Errard}, {Fabbian}, {Feeney}, {Ferraro}, {Flux{\`a}}, {Freese}, {Frisch}, {Frolov}, {Fuller}, {Fuzia}, {Galitzki}, {Gallardo}, {Tomas Galvez Ghersi}, {Gao}, {Gawiser}, {Gerbino}, {Gluscevic}, {Goeckner-Wald}, {Golec}, {Gordon}, {Gralla}, {Green}, {Grigorian}, {Groh}, {Groppi}, {Guan}, {Gudmundsson}, {Han}, {Hargrave}, {Hasegawa}, {Hasselfield}, {Hattori}, {Haynes}, {Hazumi}, {He}, {Healy}, {Henderson}, {Hervias-Caimapo}, {Hill}, {Hill}, {Hilton}, {Hilton}, {Hincks}, {Hinshaw}, {Hlo{\v{z}}ek}, {Ho}, {Ho}, {Howe}, {Huang}, {Hubmayr}, {Huffenberger}, {Hughes}, {Ijjas}, {Ikape}, {Irwin}, {Jaffe}, {Jain}, {Jeong}, {Kaneko}, {Karpel}, {Katayama}, {Keating}, {Kernasovskiy}, {Keskitalo}, {Kisner}, {Kiuchi}, {Klein}, {Knowles}, {Koopman}, {Kosowsky}, {Krachmalnicoff}, {Kuenstner}, {Kuo}, {Kusaka}, {Lashner}, {Lee}, {Lee}, {Leon}, {Leung}, {Lewis}, {Li}, {Li}, {Limon}, {Linder}, {Lopez-Caraballo}, {Louis}, {Lowry}, {Lungu}, {Madhavacheril}, {Mak}, {Maldonado}, {Mani}, {Mates}, {Matsuda}, {Maurin}, {Mauskopf}, {May}, {McCallum}, {McKenney}, {McMahon}, {Meerburg}, {Meyers}, {Miller}, {Mirmelstein}, {Moodley}, {Munchmeyer}, {Munson}, {Naess}, {Nati}, {Navaroli}, {Newburgh}, {Nguyen}, {Niemack}, {Nishino}, {Orlowski-Scherer}, {Page}, {Partridge}, {Peloton}, {Perrotta}, {Piccirillo}, {Pisano}, {Poletti}, {Puddu}, {Puglisi}, {Raum}, {Reichardt}, {Remazeilles}, {Rephaeli}, {Riechers}, {Rojas}, {Roy}, {Sadeh}, {Sakurai}, {Salatino}, {Sathyanarayana Rao}, {Schaan}, {Schmittfull}, {Sehgal}, {Seibert}, {Seljak}, {Sherwin}, {Shimon}, {Sierra}, {Sievers}, {Sikhosana}, {Silva-Feaver}, {Simon}, {Sinclair}, {Siritanasak}, {Smith}, {Smith}, {Spergel}, {Staggs}, {Stein}, {Stevens}, {Stompor}, {Suzuki}, {Tajima}, {Takakura}, {Teply}, {Thomas}, {Thorne}, {Thornton}, {Trac}, {Tsai}, {Tucker}, {Ullom}, {Vagnozzi}, {van Engelen}, {Van Lanen}, {Van Winkle}, {Vavagiakis}, {Verg{\`e}s}, {Vissers}, {Wagoner}, {Walker}, {Ward}, {Westbrook}, {Whitehorn}, {Williams}, {Williams}, {Wollack}, {Xu}, {Yu}, {Yu}, {Zago}, {Zhang}, {Zhu}, \& {Simons Observatory Collaboration}}]{SO2019}
{Ade}, P., {Aguirre}, J., {Ahmed}, Z., {et~al.} 2019, \jcap, 2019, 056

\bibitem[{{Ade} {et~al.}(2014){Ade}, {Aikin}, {Barkats}, {Benton}, {Bischoff}, {Bock}, {Brevik}, {Buder}, {Bullock}, {Dowell}, {Duband}, {Filippini}, {Fliescher}, {Golwala}, {Halpern}, {Hasselfield}, {Hildebrandt}, {Hilton}, {Hristov}, {Irwin}, {Karkare}, {Kaufman}, {Keating}, {Kernasovskiy}, {Kovac}, {Kuo}, {Leitch}, {Lueker}, {Mason}, {Netterfield}, {Nguyen}, {O'Brient}, {Ogburn}, {Orlando}, {Pryke}, {Reintsema}, {Richter}, {Schwarz}, {Sheehy}, {Staniszewski}, {Sudiwala}, {Teply}, {Tolan}, {Turner}, {Vieregg}, {Wong}, {Yoon}, \& {Bicep2 Collaboration}}]{bicep2_2014}
{Ade}, P.~A.~R., {Aikin}, R.~W., {Barkats}, D., {et~al.} 2014, Physical Review Letters, 112, 241101

\bibitem[{{Bennett} {et~al.}(2013){Bennett}, {Larson}, {Weiland}, {Jarosik}, {Hinshaw}, {Odegard}, {Smith}, {Hill}, {Gold}, {Halpern}, {Komatsu}, {Nolta}, {Page}, {Spergel}, {Wollack}, {Dunkley}, {Kogut}, {Limon}, {Meyer}, {Tucker}, \& {Wright}}]{bennett:2013}
{Bennett}, C.~L., {Larson}, D., {Weiland}, J.~L., {et~al.} 2013, \apjs, 208, 20

\bibitem[{{BeyondPlanck Collaboration}(2023)}]{bp01}
{BeyondPlanck Collaboration}. 2023, A\&A, 675, A1

\bibitem[{{BICEP2 Collaboration}(2018)}]{Bicep2018limit}
{BICEP2 Collaboration}. 2018, \prl, 121, 221301

\bibitem[{{Cutri} {et~al.}(2013){Cutri}, {Wright}, {Conrow}, {Fowler}, {Eisenhardt}, {Grillmair}, {Kirkpatrick}, {Masci}, {McCallon}, {Wheelock}, {Fajardo-Acosta}, {Yan}, {Benford}, {Harbut}, {Jarrett}, {Lake}, {Leisawitz}, {Ressler}, {Stanford}, {Tsai}, {Liu}, {Helou}, {Mainzer}, {Gettings}, {Gonzalez}, {Hoffman}, {Marsh}, {Padgett}, {Skrutskie}, {Beck}, {Papin}, \& {Wittman}}]{allwise_ES}
{Cutri}, R.~M., {Wright}, E.~L., {Conrow}, T., {et~al.} 2013, {Explanatory Supplement to the AllWISE Data Release Products}, Explanatory Supplement to the AllWISE Data Release Products, by R. M. Cutri et al.

\bibitem[{{Dame} {et~al.}(2001){Dame}, {Hartmann}, \& {Thaddeus}}]{dame2001}
{Dame}, T.~M., {Hartmann}, D., \& {Thaddeus}, P. 2001, \apj, 547, 792

\bibitem[{{D{\'\i}az-Santos} {et~al.}(2017){D{\'\i}az-Santos}, {Armus}, {Charmandaris}, {Lu}, {Stierwalt}, {Stacey}, {Malhotra}, {van der Werf}, {Howell}, {Privon}, {Mazzarella}, {Goldsmith}, {Murphy}, {Barcos-Mu{\~n}oz}, {Linden}, {Inami}, {Larson}, {Evans}, {Appleton}, {Iwasawa}, {Lord}, {Sanders}, \& {Surace}}]{2017Herschel_FIR_CII}
{D{\'\i}az-Santos}, T., {Armus}, L., {Charmandaris}, V., {et~al.} 2017, \apj, 846, 32

\bibitem[{{Edenhofer} {et~al.}(2024){Edenhofer}, {Zucker}, {Frank}, {Saydjari}, {Speagle}, {Finkbeiner}, \& {En{\ss}lin}}]{edenhofer:2024}
{Edenhofer}, G., {Zucker}, C., {Frank}, P., {et~al.} 2024, \aap, 685, A82

\bibitem[{{Eriksen} {et~al.}(2008){Eriksen}, {Jewell}, {Dickinson}, {Banday}, {G{\'o}rski}, \& {Lawrence}}]{eriksen2008}
{Eriksen}, H.~K., {Jewell}, J.~B., {Dickinson}, C., {et~al.} 2008, \apj, 676, 10

\bibitem[{{Eriksen} {et~al.}(2004){Eriksen}, {O'Dwyer}, {Jewell}, {Wand elt}, {Larson}, {G{\'o}rski}, {Levin}, {Banday}, \& {Lilje}}]{eriksen:2004}
{Eriksen}, H.~K., {O'Dwyer}, I.~J., {Jewell}, J.~B., {et~al.} 2004, \apjs, 155, 227

\bibitem[{{Finkbeiner} {et~al.}(1999){Finkbeiner}, {Davis}, \& {Schlegel}}]{finkbeiner1999}
{Finkbeiner}, D.~P., {Davis}, M., \& {Schlegel}, D.~J. 1999, \apj, 524, 867

\bibitem[{{Fixsen} {et~al.}(1998){Fixsen}, {Bennett}, \& {Mather}}]{fixsen:1998}
{Fixsen}, D.~J., {Bennett}, C.~L., \& {Mather}, J.~C. 1998, arXiv e-prints, astro

\bibitem[{{Fixsen} {et~al.}(1997){Fixsen}, {Weiland}, {Brodd}, {Hauser}, {Kelsall}, {Leisawitz}, {Mather}, {Jensen}, {Schafer}, \& {Silverberg}}]{fixsen1997}
{Fixsen}, D.~J., {Weiland}, J.~L., {Brodd}, S., {et~al.} 1997, \apj, 490, 482

\bibitem[{{Gaia Collaboration et al.}(2016)}]{gaia:2016}
{Gaia Collaboration et al.} 2016, \aap, 595, A1

\bibitem[{{Galloway} {et~al.}(2023){Galloway}, {Andersen, K. J.}, {Aurlien, R.}, {Banerji, R.}, {Bersanelli, M.}, {Bertocco, S.}, {Brilenkov, M.}, {Carbone, M.}, {Colombo, L. P. L.}, {Eriksen, H. K.}, {Eskilt, J. R.}, {Foss, M. K.}, {Franceschet, C.}, {Fuskeland, U.}, {Galeotta, S.}, {Gerakakis, S.}, {Gjerl\o{}w, E.}, {Hensley, B.}, {Herman, D.}, {Iacobellis, M.}, {Ieronymaki, M.}, {Ihle, H. T.}, {Jewell, J. B.}, {Karakci, A.}, {Keih\"anen, E.}, {Keskitalo, R.}, {Maggio, G.}, {Maino, D.}, {Maris, M.}, {Mennella, A.}, {Paradiso, S.}, {Partridge, B.}, {Reinecke, M.}, {San, M.}, {Suur-Uski, A.-S.}, {Svalheim, T. L.}, {Tavagnacco, D.}, {Thommesen, H.}, {Watts, D. J.}, {Wehus, I. K.}, \& {Zacchei, A.}}]{bp03}
{Galloway}, M., {Andersen, K. J.}, {Aurlien, R.}, {et~al.} 2023, A\&A, 675, A3

\bibitem[{{Galloway} {et~al.}(2026)}]{CG02_04}
{Galloway}, M. {et~al.} 2026, \aap, in preparation [\eprint[arXiv]{2601.07831}]

\bibitem[{Geman \& Geman(1984)}]{geman:1984}
Geman, S. \& Geman, D. 1984, IEEE Trans. Pattern Anal. Mach. Intell., 6, 721

\bibitem[{{Gjerløw et al.}(2026{\natexlab{a}})}]{CG02_05}
{Gjerløw et al.} 2026{\natexlab{a}}, \aap, in preparation [\eprint[arXiv]{2601.07818}]

\bibitem[{{Gjerløw et al.}(2026{\natexlab{b}})}]{CG02_07}
{Gjerløw et al.} 2026{\natexlab{b}}, \aap, in preparation [\eprint[arXiv]{2601.07822}]

\bibitem[{{G{\'o}rski} {et~al.}(2005){G{\'o}rski}, {Hivon}, {Banday}, {Wandelt}, {Hansen}, {Reinecke}, \& {Bartelmann}}]{healpix}
{G{\'o}rski}, K.~M., {Hivon}, E., {Banday}, A.~J., {et~al.} 2005, \apj, 622, 759

\bibitem[{{Haffner} {et~al.}(2016){Haffner}, {Reynolds}, {Babler}, {Madsen}, {Hill}, {Barger}, {Jaehnig}, {Mierkiewicz}, {Percival}, {Chopra}, {Pingel}, {Reese}, {Gostisha}, \& {Wunderlin}}]{2016WHAM}
{Haffner}, L.~M., {Reynolds}, R.~J., {Babler}, B.~L., {et~al.} 2016, in American Astronomical Society Meeting Abstracts, Vol. 227, American Astronomical Society Meeting Abstracts \#227, 347.17

\bibitem[{{Haffner} {et~al.}(2003){Haffner}, {Reynolds}, {Tufte}, {Madsen}, {Jaehnig}, \& {Percival}}]{wham:2003}
{Haffner}, L.~M., {Reynolds}, R.~J., {Tufte}, S.~L., {et~al.} 2003, \apjs, 149, 405

\bibitem[{{Haslam} {et~al.}(1982){Haslam}, {Salter}, {Stoffel}, \& {Wilson}}]{haslam1982}
{Haslam}, C.~G.~T., {Salter}, C.~J., {Stoffel}, H., \& {Wilson}, W.~E. 1982, \aaps, 47, 1

\bibitem[{{Hauser} {et~al.}(1998){Hauser}, {Arendt}, {Kelsall}, {Dwek}, {Odegard}, {Weiland}, {Freudenreich}, {Reach}, {Silverberg}, {Moseley}, {Pei}, {Lubin}, {Mather}, {Shafer}, {Smoot}, {Weiss}, {Wilkinson}, \& {Wright}}]{hauser1998}
{Hauser}, M.~G., {Arendt}, R.~G., {Kelsall}, T., {et~al.} 1998, \apj, 508, 25

\bibitem[{{Hensley} \& {Draine}(2021)}]{Hensley2021}
{Hensley}, B.~S. \& {Draine}, B.~T. 2021, \apj, 906, 73

\bibitem[{{Hensley} \& {Draine}(2023)}]{Hensley2023}
{Hensley}, B.~S. \& {Draine}, B.~T. 2023, \apj, 948, 55

\bibitem[{{HI4PI Collaboration} {et~al.}(2016){HI4PI Collaboration}, {Ben Bekhti}, {Fl{\"o}er}, {Keller}, {Kerp}, {Lenz}, {Winkel}, {Bailin}, {Calabretta}, {Dedes}, {Ford}, {Gibson}, {Haud}, {Janowiecki}, {Kalberla}, {Lockman}, {McClure-Griffiths}, {Murphy}, {Nakanishi}, {Pisano}, \& {Staveley-Smith}}]{HI4PI2016}
{HI4PI Collaboration}, {Ben Bekhti}, N., {Fl{\"o}er}, L., {et~al.} 2016, \aap, 594, A116

\bibitem[{{Hocuk} {et~al.}(2017){Hocuk}, {Sz{\H{u}}cs}, {Caselli}, {Cazaux}, {Spaans}, \& {Esplugues}}]{2017A&A_DustParam}
{Hocuk}, S., {Sz{\H{u}}cs}, L., {Caselli}, P., {et~al.} 2017, \aap, 604, A58

\bibitem[{{Jackson} {et~al.}(2020){Jackson}, {Allingham}, {Killerby-Smith}, {Whitaker}, {Smith}, {Contreras}, {Guzm{\'a}n}, {Hogge}, {Sanhueza}, \& {Stephens}}]{2020CII}
{Jackson}, J.~M., {Allingham}, D., {Killerby-Smith}, N., {et~al.} 2020, \apj, 904, 18

\bibitem[{{Juvela} \& {Ysard}(2012)}]{juvela:2012}
{Juvela}, M. \& {Ysard}, N. 2012, \aap, 541, A33

\bibitem[{{Kelsall} {et~al.}(1998){Kelsall}, {Weiland}, {Franz}, {Reach}, {Arendt}, {Dwek}, {Freudenreich}, {Hauser}, {Moseley}, {Odegard}, {Silverberg}, \& {Wright}}]{K98}
{Kelsall}, T., {Weiland}, J.~L., {Franz}, B.~A., {et~al.} 1998, \apj, 508, 44

\bibitem[{{Klessen} \& {Glover}(2016)}]{2016SAA_DustReview}
{Klessen}, R.~S. \& {Glover}, S. C.~O. 2016, Saas-Fee Advanced Course, 43, 85

\bibitem[{{Lallement} {et~al.}(2022){Lallement}, {Vergely}, {Babusiaux}, \& {Cox}}]{lallement:2022}
{Lallement}, R., {Vergely}, J.~L., {Babusiaux}, C., \& {Cox}, N.~L.~J. 2022, \aap, 661, A147

\bibitem[{{Langer} {et~al.}(2014){Langer}, {Pineda}, \& {Velusamy}}]{2014_CII}
{Langer}, W.~D., {Pineda}, J.~L., \& {Velusamy}, T. 2014, \aap, 564, A101

\bibitem[{{Langer} {et~al.}(2010){Langer}, {Velusamy}, {Pineda}, {Goldsmith}, {Li}, \& {Yorke}}]{2010_CII}
{Langer}, W.~D., {Velusamy}, T., {Pineda}, J.~L., {et~al.} 2010, \aap, 521, L17

\bibitem[{{Lenz} {et~al.}(2017){Lenz}, {Hensley}, \& {Dor{\'e}}}]{lenz:2017}
{Lenz}, D., {Hensley}, B.~S., \& {Dor{\'e}}, O. 2017, \apj, 846, 38

\bibitem[{{LiteBIRD Collaboration} {et~al.}(2023){LiteBIRD Collaboration}, {Allys}, {Arnold}, {Aumont}, {Aurlien}, {Azzoni}, {Baccigalupi}, {Banday}, {Banerji}, {Barreiro}, {Bartolo}, {Bautista}, {Beck}, {Beckman}, {Bersanelli}, {Boulanger}, {Brilenkov}, {Bucher}, {Calabrese}, {Campeti}, {Carones}, {Casas}, {Catalano}, {Chan}, {Cheung}, {Chinone}, {Clark}, {Columbro}, {D'Alessandro}, {de Bernardis}, {de Haan}, {de la Hoz}, {De Petris}, {Torre}, {Diego-Palazuelos}, {Dobbs}, {Dotani}, {Duval}, {Elleflot}, {Eriksen}, {Errard}, {Essinger-Hileman}, {Finelli}, {Flauger}, {Franceschet}, {Fuskeland}, {Galloway}, {Ganga}, {Gerbino}, {Gervasi}, {G{\'e}nova-Santos}, {Ghigna}, {Giardiello}, {Gjerl{\o}w}, {Grain}, {Grupp}, {Gruppuso}, {Gudmundsson}, {Halverson}, {Hargrave}, {Hasebe}, {Hasegawa}, {Hazumi}, {Henrot-Versill{\'e}}, {Hensley}, {Hergt}, {Herman}, {Hivon}, {Hlozek}, {Hornsby}, {Hoshino}, {Hubmayr}, {Ichiki}, {Iida}, {Imada}, {Ishino}, {Jaehnig}, {Katayama}, {Kato}, {Keskitalo}, {Kisner}, {Kobayashi}, {Kogut}, {Kohri}, {Komatsu}, {Komatsu}, {Konishi}, {Krachmalnicoff}, {Kuo}, {Lamagna}, {Lattanzi}, {Lee}, {Leloup}, {Levrier}, {Linder}, {Luzzi}, {Macias-Perez}, {Maciaszek}, {Maffei}, {Maino}, {Mandelli}, {Mart{\'\i}nez-Gonz{\'a}lez}, {Masi}, {Massa}, {Matarrese}, {Matsuda}, {Matsumura}, {Mele}, {Migliaccio}, {Minami}, {Moggi}, {Montgomery}, {Montier}, {Morgante}, {Mot}, {Nagano}, {Nagasaki}, {Nagata}, {Nakano}, {Namikawa}, {Nati}, {Natoli}, {Nerval}, {Noviello}, {Odagiri}, {Oguri}, {Ohsaki}, {Pagano}, {Paiella}, {Paoletti}, {Passerini}, {Patanchon}, {Piacentini}, {Piat}, {Pisano}, {Polenta}, {Poletti}, {Prouv{\'e}}, {Puglisi}, {Rambaud}, {Raum}, {Realini}, {Reinecke}, {Remazeilles}, {Ritacco}, {Roudil}, {Rubino-Martin}, {Russell}, {Sakurai}, {Sakurai}, {Sasaki}, {Scott}, {Sekimoto}, {Shinozaki}, {Shiraishi}, {Shirron}, {Signorelli}, {Spinella}, {Stever}, {Stompor}, {Sugiyama}, {Sullivan}, {Suzuki}, {Svalheim}, {Switzer}, {Takaku}, {Takakura}, {Takase}, {Tartari}, {Terao}, {Thermeau}, {Thommesen}, {Thompson}, {Tomasi}, {Tominaga}, {Tristram}, {Tsuji}, {Tsujimoto}, {Vacher}, {Vielva}, {Vittorio}, {Wang}, {Watanuki}, {Wehus}, {Weller}, {Westbrook}, {Wilms}, {Winter}, {Wollack}, {Yumoto}, {Zannoni}, \& {Collaboration LiteB I R D}}]{litebird2022}
{LiteBIRD Collaboration}, {Allys}, E., {Arnold}, K., {et~al.} 2023, Progress of Theoretical and Experimental Physics, 2023, 042F01

\bibitem[{{Maris} {et~al.}(2006){Maris}, {Burigana}, \& {Fogliani}}]{maris2006c}
{Maris}, M., {Burigana}, C., \& {Fogliani}, S. 2006, \aap, 452, 685

\bibitem[{{Mather} {et~al.}(1994){Mather}, {Cheng}, {Cottingham}, {Eplee}, {Fixsen}, {Hewagama}, {Isaacman}, {Jensen}, {Meyer}, {Noerdlinger}, {Read}, {Rosen}, {Shafer}, {Wright}, {Bennett}, {Boggess}, {Hauser}, {Kelsall}, {Moseley}, {Silverberg}, {Smoot}, {Weiss}, \& {Wilkinson}}]{mather:1994}
{Mather}, J.~C., {Cheng}, E.~S., {Cottingham}, D.~A., {et~al.} 1994, \apj, 420, 439

\bibitem[{{Montegriffo} {et~al.}(2023){Montegriffo}, {De Angeli}, {Andrae}, {Riello}, {Pancino}, {Sanna}, {Bellazzini}, {Evans}, {Carrasco}, {Sordo}, {Busso}, {Cacciari}, {Jordi}, {van Leeuwen}, {Vallenari}, {Altavilla}, {Barstow}, {Brown}, {Burgess}, {Castellani}, {Cowell}, {Davidson}, {De Luise}, {Delchambre}, {Diener}, {Fabricius}, {Fr{\'e}mat}, {Fouesneau}, {Gilmore}, {Giuffrida}, {Hambly}, {Harrison}, {Hidalgo}, {Hodgkin}, {Holland}, {Marinoni}, {Osborne}, {Pagani}, {Palaversa}, {Piersimoni}, {Pulone}, {Ragaini}, {Rainer}, {Richards}, {Rowell}, {Ruz-Mieres}, {Sarro}, {Walton}, \& {Yoldas}}]{2023gaia}
{Montegriffo}, P., {De Angeli}, F., {Andrae}, R., {et~al.} 2023, \aap, 674, A3

\bibitem[{Notari \& Quartin(2015)}]{Notari:2015}
Notari, A. \& Quartin, M. 2015, Journal of Cosmology and Astroparticle Physics, 2015, 047–047

\bibitem[{{Partridge} \& {Peebles}(1967)}]{partridge1967}
{Partridge}, R.~B. \& {Peebles}, P.~J.~E. 1967, \apj, 148, 377

\bibitem[{{Penzias} \& {Wilson}(1965)}]{penzias:1965}
{Penzias}, A.~A. \& {Wilson}, R.~W. 1965, \apj, 142, 419

\bibitem[{{Pineda} {et~al.}(2013){Pineda}, {Langer}, {Velusamy}, \& {Goldsmith}}]{2013_CII}
{Pineda}, J.~L., {Langer}, W.~D., {Velusamy}, T., \& {Goldsmith}, P.~F. 2013, \aap, 554, A103

\bibitem[{{Planck Collaboration}(2018)}]{Planck2018xii}
{Planck Collaboration}. 2018, arXiv e-prints, arXiv:1807.06212

\bibitem[{{\sorthelp{Planck Collaboration 2011J}}{Planck Collaboration X}(2011)}]{planck2011-5.2a}
{\sorthelp{Planck Collaboration 2011J}}{Planck Collaboration X}. 2011, \aap, 536, A10

\bibitem[{{\sorthelp{Planck Collaboration 2014F}}{Planck Collaboration VI}(2014)}]{planck2013-p03}
{\sorthelp{Planck Collaboration 2014F}}{Planck Collaboration VI}. 2014, \aap, 571, A6

\bibitem[{{\sorthelp{Planck Collaboration 2014K}}{Planck Collaboration XI}(2014)}]{planck2013-p06b}
{\sorthelp{Planck Collaboration 2014K}}{Planck Collaboration XI}. 2014, \aap, 571, A11

\bibitem[{{\sorthelp{Planck Collaboration 2014N}}{Planck Collaboration XIV}(2014)}]{planck2013-pip88}
{\sorthelp{Planck Collaboration 2014N}}{Planck Collaboration XIV}. 2014, \aap, 571, A14

\bibitem[{{\sorthelp{Planck Collaboration 2015H}}{Planck Collaboration VIII}(2016)}]{planck2014-a09}
{\sorthelp{Planck Collaboration 2015H}}{Planck Collaboration VIII}. 2016, \aap, 594, A8

\bibitem[{{\sorthelp{Planck Collaboration 2015J}}{Planck Collaboration X}(2016)}]{planck2014-a12}
{\sorthelp{Planck Collaboration 2015J}}{Planck Collaboration X}. 2016, \aap, 594, A10

\bibitem[{{\sorthelp{Planck Collaboration 2018A}}{Planck Collaboration I}(2020)}]{planck2016-l01}
{\sorthelp{Planck Collaboration 2018A}}{Planck Collaboration I}. 2020, \aap, 641, A1

\bibitem[{{\sorthelp{Planck Collaboration 2018C}}{Planck Collaboration III}(2020)}]{planck2016-l03}
{\sorthelp{Planck Collaboration 2018C}}{Planck Collaboration III}. 2020, \aap, 641, A3

\bibitem[{{\sorthelp{Planck Collaboration 2018D}}{Planck Collaboration IV}(2020)}]{planck2016-l04}
{\sorthelp{Planck Collaboration 2018D}}{Planck Collaboration IV}. 2020, \aap, 641, A4

\bibitem[{{\sorthelp{Planck Collaboration IntY}}{Planck Collaboration Int. XXV}(2015)}]{planck2014-XXV}
{\sorthelp{Planck Collaboration IntY}}{Planck Collaboration Int. XXV}. 2015, \aap, 582, A28

\bibitem[{{\sorthelp{Planck Collaboration IntZA}}{Planck Collaboration Int. XXVI}(2015)}]{planck2014-XXVI}
{\sorthelp{Planck Collaboration IntZA}}{Planck Collaboration Int. XXVI}. 2015, \aap, 582, A29

\bibitem[{{\sorthelp{Planck Collaboration IntZW}}{Planck Collaboration Int. XLVIII}(2016)}]{planck2016-XLVIII}
{\sorthelp{Planck Collaboration IntZW}}{Planck Collaboration Int. XLVIII}. 2016, \aap, 596, A109

\bibitem[{{\sorthelp{Planck Collaboration IntZZF}}{Planck Collaboration Int. LVI}(2020)}]{planck2020-LVI}
{\sorthelp{Planck Collaboration IntZZF}}{Planck Collaboration Int. LVI}. 2020, \aap, 644, 100

\bibitem[{{\sorthelp{Planck Collaboration IntZZG}}{Planck Collaboration Int. LVII}(2020)}]{planck2020-LVII}
{\sorthelp{Planck Collaboration IntZZG}}{Planck Collaboration Int. LVII}. 2020, \aap, 643, 42

\bibitem[{{San} {et~al.}(2024)}]{CG02_02}
{San}, M. {et~al.} 2024, \aap, in preparation [\eprint[arXiv]{20xx.xxxxx}]

\bibitem[{{Seljebotn} {et~al.}(2019){Seljebotn}, {B{\ae}rland}, {Eriksen}, {Mardal}, \& {Wehus}}]{seljebotn:2019}
{Seljebotn}, D.~S., {B{\ae}rland}, T., {Eriksen}, H.~K., {Mardal}, K.~A., \& {Wehus}, I.~K. 2019, \aap, 627, A98

\bibitem[{{Skrutskie} {et~al.}(2006){Skrutskie}, {Cutri}, {Stiening}, {Weinberg}, {Schneider}, {Carpenter}, {Beichman}, {Capps}, {Chester}, {Elias}, {Huchra}, {Liebert}, {Lonsdale}, {Monet}, {Price}, {Seitzer}, {Jarrett}, {Kirkpatrick}, {Gizis}, {Howard}, {Evans}, {Fowler}, {Fullmer}, {Hurt}, {Light}, {Kopan}, {Marsh}, {McCallon}, {Tam}, {Van Dyk}, \& {Wheelock}}]{2006AJ_2mass}
{Skrutskie}, M.~F., {Cutri}, R.~M., {Stiening}, R., {et~al.} 2006, \aj, 131, 1163

\bibitem[{{Wang} {et~al.}(2022{\natexlab{a}}){Wang}, {Huang}, {Yuan}, {Zhang}, {Xiang}, \& {Liu}}]{2022ApJS_lamost}
{Wang}, C., {Huang}, Y., {Yuan}, H., {et~al.} 2022{\natexlab{a}}, \apjs, 259, 51

\bibitem[{{Wang} {et~al.}(2022{\natexlab{b}}){Wang}, {Li}, {Wu}, {Gies}, {Liu}, {Liu}, {Guo}, {Chen}, \& {Han}}]{2022yCat_lamost}
{Wang}, L., {Li}, J., {Wu}, Y., {et~al.} 2022{\natexlab{b}}, {VizieR Online Data Catalog: Classical Be stars from LAMOST MRS DR7 (Wang+, 2022)}, VizieR On-line Data Catalog: J/ApJS/260/35. Originally published in: 2022ApJS..260...35W

\bibitem[{{Watts} {et~al.}(2024{\natexlab{a}}){Watts}, {Galloway}, {Gjerlow}, {San}, {Aurlien}, {et~al.}}]{CG02_01}
{Watts}, D., {Galloway}, M., {Gjerlow}, E., {et~al.} 2024{\natexlab{a}}, \aap, submitted [\eprint[arXiv]{2408.10952}]

\bibitem[{{Watts} {et~al.}(2024{\natexlab{b}})}]{CG02_03}
{Watts}, D. {et~al.} 2024{\natexlab{b}}, \aap, in preparation [\eprint[arXiv]{2406.01491}]

\bibitem[{{Wright} {et~al.}(2010){Wright}, {Eisenhardt}, {Mainzer}, {Ressler}, {Cutri}, {Jarrett}, {Kirkpatrick}, {Padgett}, {McMillan}, {Skrutskie}, {Stanford}, {Cohen}, {Walker}, {Mather}, {Leisawitz}, {Gautier}, {McLean}, {Benford}, {Lonsdale}, {Blain}, {Mendez}, {Irace}, {Duval}, {Liu}, {Royer}, {Heinrichsen}, {Howard}, {Shannon}, {Kendall}, {Walsh}, {Larsen}, {Cardon}, {Schick}, {Schwalm}, {Abid}, {Fabinsky}, {Naes}, \& {Tsai}}]{2010AJ_wise}
{Wright}, E.~L., {Eisenhardt}, P. R.~M., {Mainzer}, A.~K., {et~al.} 2010, \aj, 140, 1868

\bibitem[{{Xiang} \& {Rix}(2023)}]{2023yCatp038060301X_lamost}
{Xiang}, M. \& {Rix}, H.-W. 2023, VizieR Online Data Catalog (other), 0380, J/other/Nat/603

\bibitem[{{Xiang} {et~al.}(2022){Xiang}, {Rix}, {Ting}, {Kudritzki}, {Conroy}, {Zari}, {Shi}, {Przybilla}, {Ramirez-Tannus}, {Tkachenko}, {Gebruers}, \& {Liu}}]{2022A&A_lamost}
{Xiang}, M., {Rix}, H.-W., {Ting}, Y.-S., {et~al.} 2022, \aap, 662, A66

\bibitem[{{Zhang} {et~al.}(2023){Zhang}, {Green}, \& {Rix}}]{2023Zhang}
{Zhang}, X., {Green}, G.~M., \& {Rix}, H.-W. 2023, \mnras, 524, 1855

\bibitem[{Zonca {et~al.}(2019)Zonca, Singer, Lenz, Reinecke, Rosset, Hivon, \& Gorski}]{Zonca2019}
Zonca, A., Singer, L., Lenz, D., {et~al.} 2019, Journal of Open Source Software, 4, 1298

\bibitem[{{Zucker} {et~al.}(2019){Zucker}, {Speagle}, {Schlafly}, {Green}, {Finkbeiner}, {Goodman}, \& {Alves}}]{2019GAIA}
{Zucker}, C., {Speagle}, J.~S., {Schlafly}, E.~F., {et~al.} 2019, \apj, 879, 125

\end{thebibliography}

\clearpage
\appendix
\onecolumn

\section{RMS scaling}
\label{app:RMSScale}
The published single horn RMS maps were found to have a known residual multiplicative offset from the production pipeline, as is evident when the single horn data maps and RMS maps are compared to each-other and to the single frequency maps (averaged maps). We include in \cref{tab:RMSScales} the values used here to divide the single horn RMS maps. These offsets were largest for the 143, 545 and 857\,GHz maps, but correspond also to a small offset to specifically the polarization sensitive bolometer maps in the case of the 100, 217 and 353\,GHz maps. 

\begin{table}[h]
    \caption{RMS scales used in this analysis. }
\begingroup
\newdimen\tblskip \tblskip=5pt
\nointerlineskip
\vskip -5mm
\footnotesize
\setbox\tablebox=\vbox{
     \newdimen\digitwidth
 \setbox0=\hbox{\rm 0}
 \digitwidth=\wd0
 \catcode`*=\active
 \def*{\kern\digitwidth}
  \newdimen\dpwidth
  \setbox0=\hbox{.}
  \dpwidth=\wd0
  \catcode`!=\active
  \def!{\kern\dpwidth}

\halign{\tabskip 0pt \hbox to 0.07\linewidth{#\leaderfil}\tabskip 0pt&\hbox to 0.07\linewidth{\hfil#\hfil}\tabskip 0pt&\hbox to 
0.07\linewidth{\hfil#\hfil}\tabskip 0pt&\hbox to 
0.07\linewidth{\hfil#\hfil}\tabskip 0pt&\hbox to 
0.07\linewidth{\hfil#\hfil}\tabskip 0pt&\hbox to 
0.07\linewidth{\hfil#\hfil}\tabskip 0pt&\hbox to 
0.07\linewidth{\hfil#\hfil}\tabskip 0pt&\hbox to 
0.07\linewidth{\hfil#\hfil}\tabskip 0pt&\hbox to 0.07\linewidth{\hfil#}\tabskip 0pt\cr
\noalign{\doubleline\vskip 1pt}
\omit\hbox to 1in{Frequency\hfil}& Horn 1  &
Horn 2  &
Horn 3  &
Horn 4  &
Horn 5  &
Horn 6  &
Horn 7  &
Horn 8 \cr
\noalign{\vskip 4pt\hrule\vskip 3pt}
\noalign{\vskip 3pt}
100 GHz& 1.414 & 1.414 & 1.414 & 1.414 & - & - & - & - \cr
\noalign{\vskip 3pt}
143 GHz& 1.85 & 2.1 & 2.03 & 1.88 & 1.68 & 1.6 & 1.75 & - \cr
\noalign{\vskip 3pt}
217 GHz& 1.414 & 1.414 &1.414  &1.414  & 1.0 & 1.0 & 1.0 & 1.0 \cr
\noalign{\vskip 3pt}
353 GHz&1.0 & 1.0 & 1.414 & 1.414 & 1.414 & 1.414 & 1.0 & 1.0 \cr
\noalign{\vskip 3pt}
545 GHz& 3.2 & 3.25 & - & 3.05 & - & - & - & - \cr
\noalign{\vskip 3pt}
857 GHz& 9.8 & 10.6 & 10.3 & 8.0 & - & - & - & - \cr
\noalign{\vskip 3pt}
\noalign{\vskip 3pt\hrule\vskip 4pt}}}
\endPlancktable
\endgroup
\label{tab:RMSScales}
\end{table}

\section{Dust templates}
\label{app:dustTemplates}
\begin{figure*}[h]
    \centering
\includegraphics[width=0.49\linewidth]{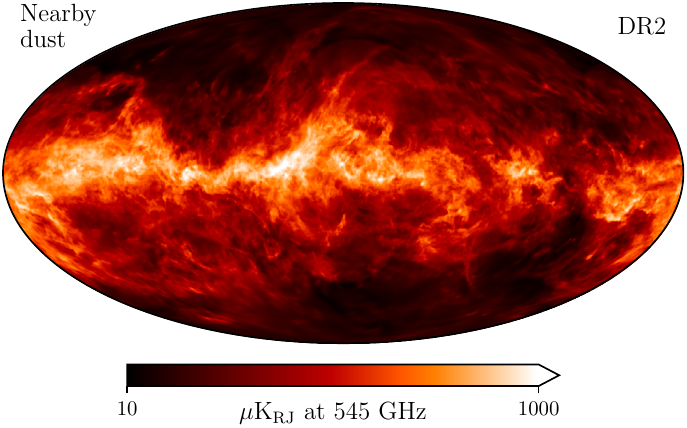}
\includegraphics[width=0.49\linewidth]{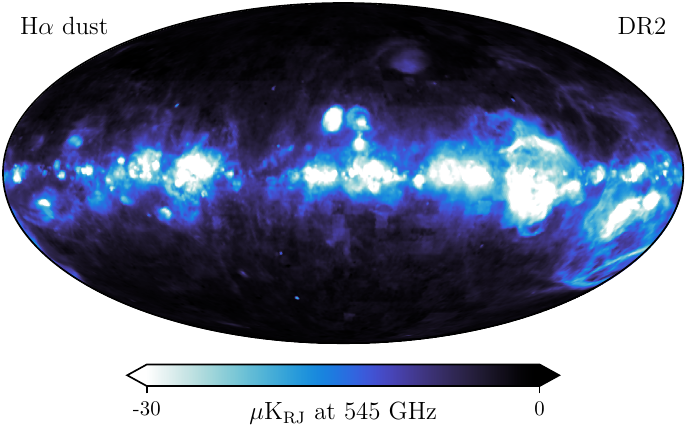}
    \caption{Nearby dust and \Ha\ dust extinction template fits.}
    \label{fig:dust_template_figures}
\end{figure*}

The templates used for the nearby dust and the \Ha\ dust are the stellar dust extinction from \cite{edenhofer:2024} and the WHAM maps \citep{wham:2003,2016WHAM}. The fits from this analysis can be seen in \cref{fig:dust_template_figures}. Note that the \Ha\ dust is negative as it represents a dust extinction component for the HFI bands, reducing the apparent dust in the blue areas. The nearby dust is mapped out to 1.25~kpc from the Sun, and has a higher amplitude away from the Galactic plane when compared to the other dust components (see also \cref{app:dustCharacterization}). 

\clearpage
\section{Residuals}
\label{app:residuals}

\noindent\begin{minipage}{0.83\textwidth}
\vspace*{1mm}
\centering
\hspace*{1.5cm}\includegraphics[width=\textwidth]{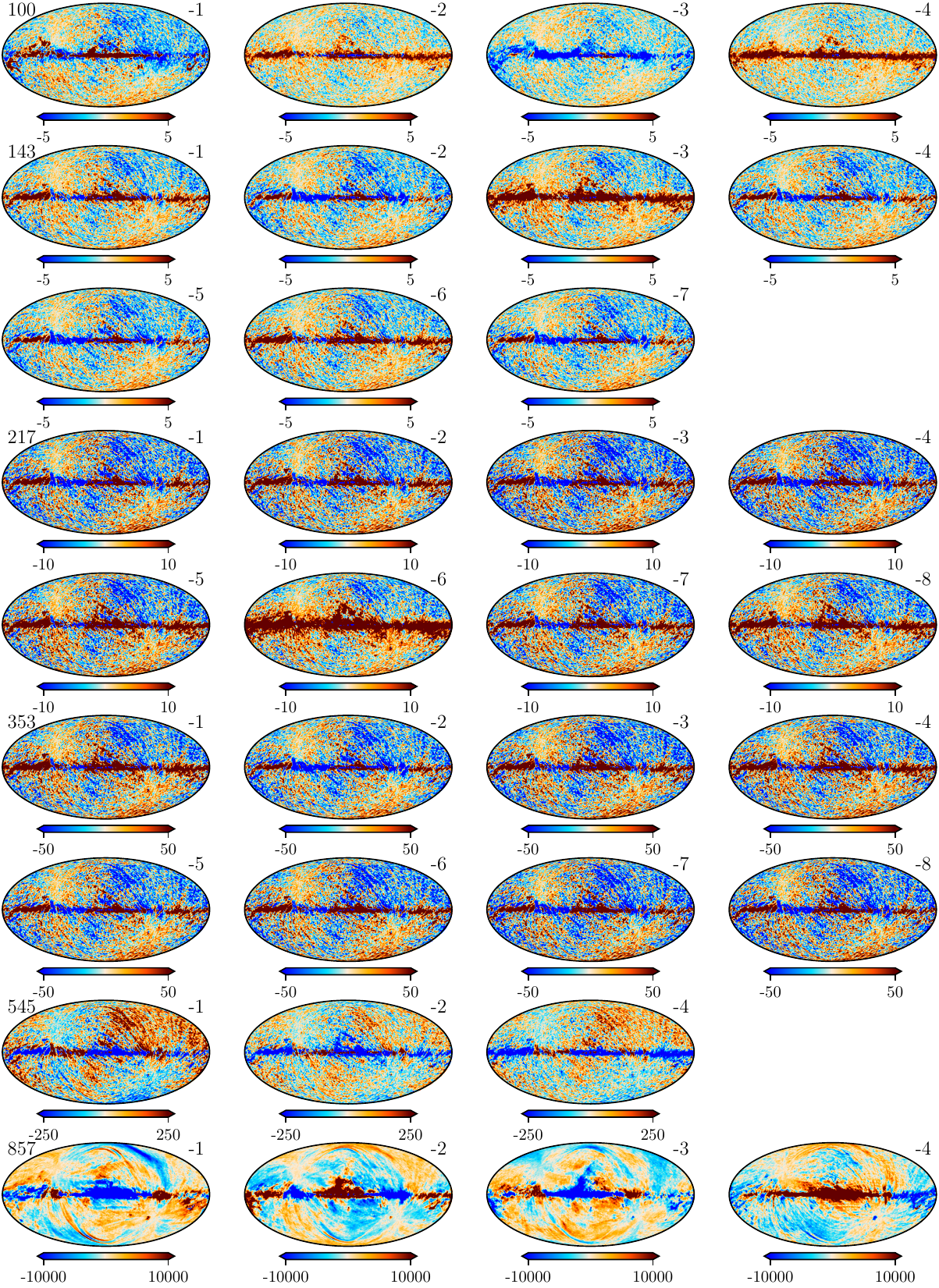}
\end{minipage}
\captionof{figure}{Residuals (Data minus model) for the \planck\ HFI bands. Units are in $\mu\mathrm{K_CMB}$. The residuals hint at some uncorrected zodiacal light along the ecliptic plane, as well as potentially some offsets to the gain of the 545 channels.}
\label{fig:residuals}
\vspace{5pt}

We can see the fidelity of the sky model by looking at the difference between the data and the model at each of the bands (see \cref{fig:residuals}). The residuals indicate potential for improvement to the modelling of the zodiacal light, as well as improvements to the 545\,GHz gain. The incredibly low residuals in the 100, 143, 217 and 353\,GHz maps demonstrate the success of this model.

\section{Grid search tests}
\label{app:GridSearchTests}
\vspace*{-3mm}
We plot several of the maps from the grid search in \cref{sec:results}. The \Ha\ dust and nearby dust templates do not change by eye (only scaling for each frequency given a set of $T,\beta,a$) so they are not plotted here. Thus, we only plot the amplitudes of the cold and hot dust for a subset of the grid points. In each plot the central map is the fiducial point, with surrounding plots varying the spectral index along the vertical axis and the temperature along the y-axis. 

\noindent\begin{minipage}{\textwidth}

\centering
\hspace*{1.5cm}\includegraphics[width=0.8\textwidth]{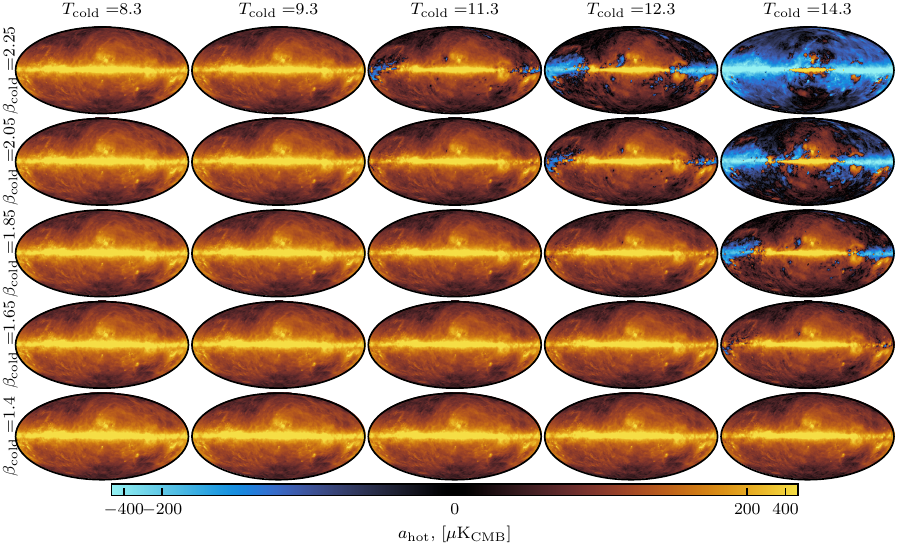}
\captionof{figure}{Hot dust amplitudes for varying $T_\mathrm{cold}$ and $\beta_\mathrm{cold}$ values with all other grid parameters set to the fiducial values as recorded in \cref{tab:SEDs}. The center panel is the fiducial map. The corresponding cold dust maps ratios can be seen in \cref{fig:cold_cold_dust_set2}.}
\label{fig:cold_hot_dust_set}

\centering
\hspace*{1.5cm}\includegraphics[width=0.8\textwidth]{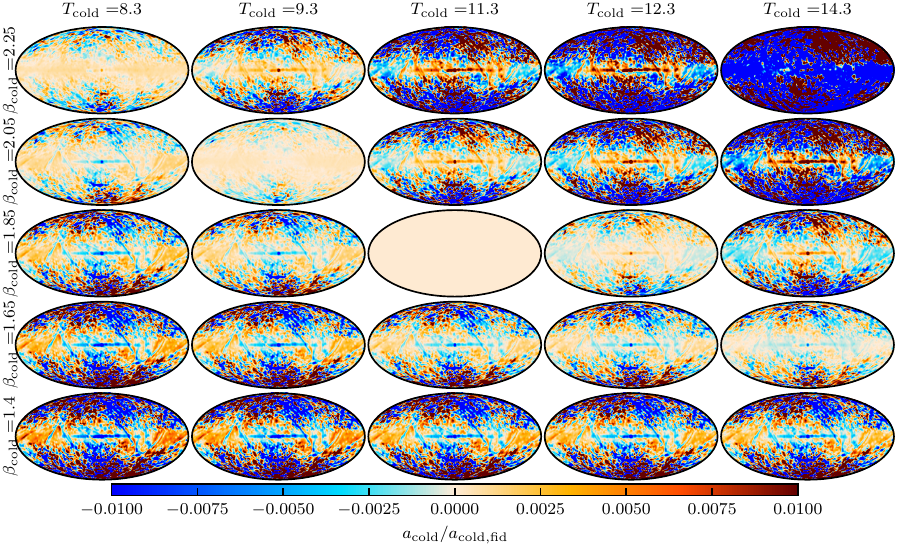}
\captionof{figure}{Monopole subtracted ratio of the cold dust amplitudes compared to the fiducial value (central panel) for varying $T_\mathrm{cold}$ and $\beta_\mathrm{cold}$. The corresponding hot dust maps can be seen in \cref{fig:cold_hot_dust_set}.}
\label{fig:cold_cold_dust_set2}
\end{minipage}

\begin{figure*}[t]
    \centering
    \includegraphics[width=0.85\linewidth]{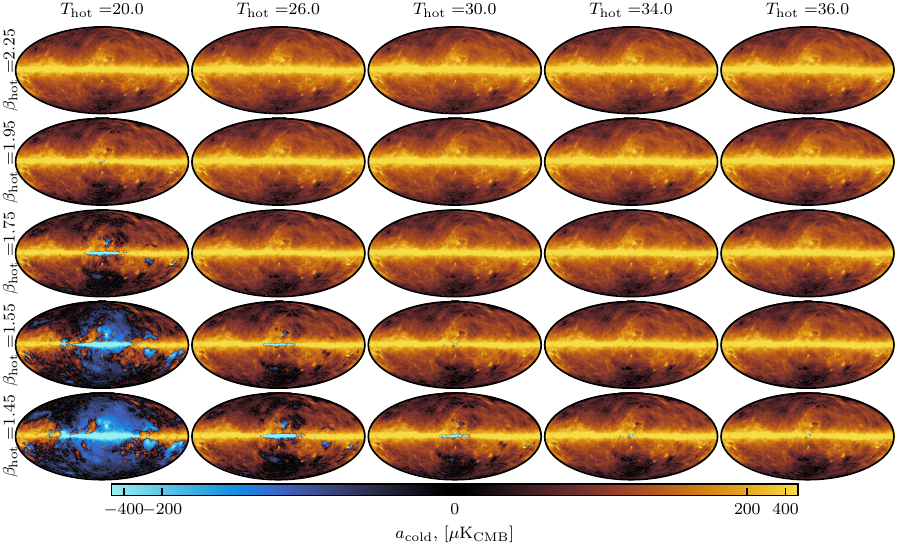}
    \caption{Cold dust amplitudes for varying $T_\mathrm{hot}$ and $\beta_\mathrm{hot}$ values with all other grid parameters set to the fiducial values as recorded in \cref{tab:SEDs}. The central panel is the fiducial map. The corresponding hot dust maps ratios can be seen in \cref{fig:hot_hot_dust_ratio}.}
    \label{fig:hot_cold_dust_set}

    \centering
    \includegraphics[width=0.85\linewidth]{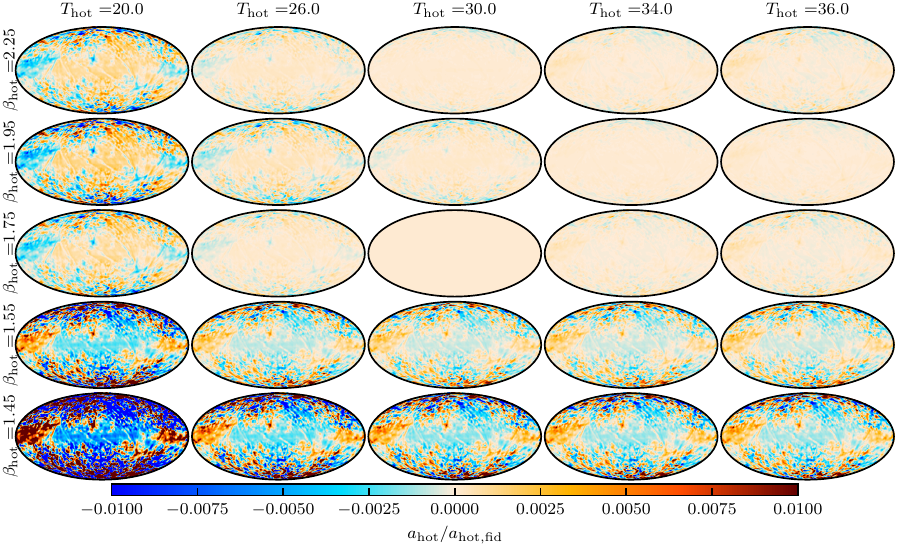}
    \caption{Monopole subtracted ratio of the hot dust amplitudes compared to the fiducial value (central panel) for varying $T_\mathrm{hot}$ and $\beta_\mathrm{hot}$. The corresponding cold dust maps can be seen in \cref{fig:hot_cold_dust_set}}
    \label{fig:hot_hot_dust_ratio}    
\end{figure*}

\begin{figure*}[t]
    \centering
    \includegraphics[width=0.85\linewidth]{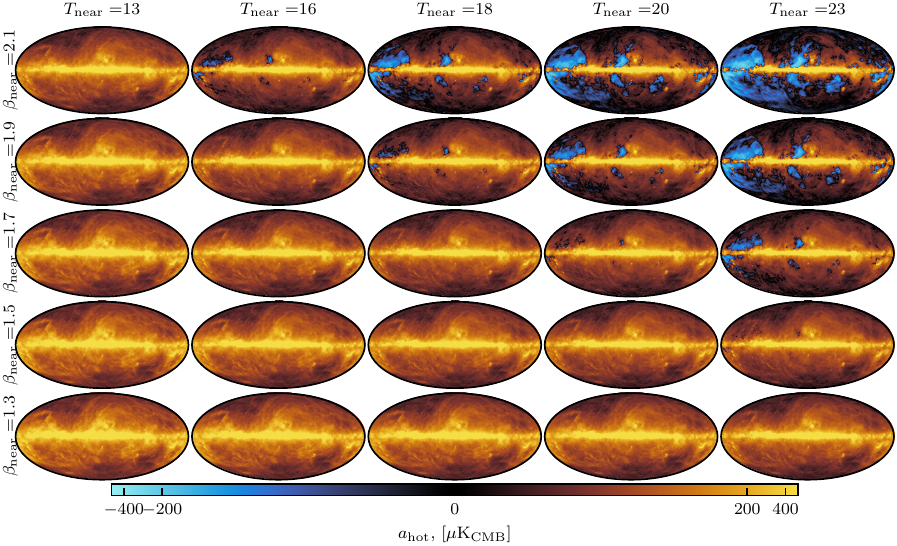}
    \caption{Hot dust amplitudes for varying $T_\mathrm{near}$ and $\beta_\mathrm{near}$ values with all other grid parameters set to the fiducial values as recorded in \cref{tab:SEDs}. The center panel is the fiducial map. The corresponding cold dust maps can be seen in \cref{fig:near_cold_dust_set}. }
    \label{fig:near_hot_dust_set}

    \centering
    \includegraphics[width=0.85\linewidth]{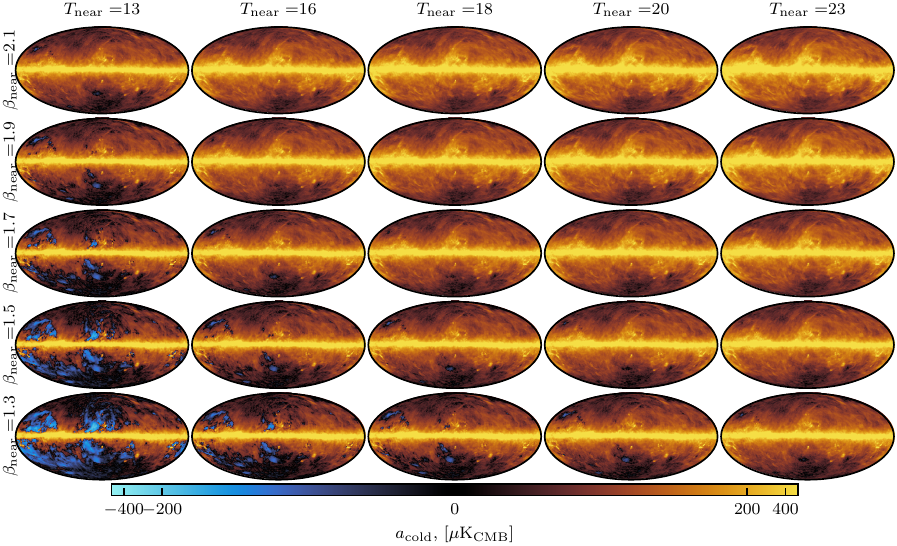}
    \caption{Cold dust amplitudes for varying $T_\mathrm{near}$ and $\beta_\mathrm{near}$ values with all other grid parameters set to the fiducial values as recorded in \cref{tab:SEDs}. The central panel is the fiducial map. The corresponding hot dust maps can be seen in \cref{fig:near_hot_dust_set}.}
    \label{fig:near_cold_dust_set}
\end{figure*}

\vspace*{-3mm}
\subsection{Cold dust}
As can be seen in \cref{fig:SEDGridContours}, there is an unphysical space mapped by the hottest and highest spectral indices in the cold dust grid. This corresponds to large over-subtraction and negative areas in the hot dust amplitudes as can be seen in \cref{fig:cold_hot_dust_set}. When shifting the $T_\mathrm{cold}$ and $\beta_\mathrm{cold}$, we find the shape of the amplitude maps is visually similar to the fiducial map (see \cref{fig:dust_figures}), so we show the monopole subtracted ratio of the grid position to the fiducial value in \cref{fig:cold_cold_dust_set2}.

\subsection{Hot dust}
Similar to with the cold dust, in \cref{fig:SEDGridContours}, there is an unphysical space mapped by the coldest and lowest spectral index in the hot dust grid. This corresponds to large over-subtraction and negative areas in the cold dust amplitudes as can be seen in \cref{fig:hot_cold_dust_set}. Similar to the cold dust, when shifting the $T_\mathrm{hot}$ and $\beta_\mathrm{hot}$ we find the shape of the amplitude maps is visually similar to the fiducial map (see \cref{fig:dust_figures}), so we show the monopole subtracted ratio of the grid position to the fiducial value in \cref{fig:hot_hot_dust_ratio}.

\subsection{Nearby Dust}
The nearby dust was found to have unphysical regions banded both by the cold dust amplitudes and the hot dust amplitudes. We additionally aimed to minimize correlations between the cold and nearby, and hot and nearby, dust, such that the dust population described by the nearby dust was an independent dust component. For regions with lower $\beta_\mathrm{near}$ and lower $T_\mathrm{near}$ the cold dust starts to behave unphysically and this corresponds also to a higher correlation between the nearby dust and the hot dust (visible in the lower left panels of \cref{fig:near_hot_dust_set}). 
Conversely, for regions with higher $\beta_\mathrm{near}$ and higher $T_\mathrm{near}$ the hot dust starts to behave unphysically and this corresponds also to a higher correlation between the nearby dust and the cold dust (as can be seen by eye in the upper right panels of \cref{fig:near_cold_dust_set}). 

\subsection{\texorpdfstring{H$\alpha$}{Ha} dust}
The parameter space explored for the \Ha\ correlated dust was shown to have a very flat $\chi^2$ profile, and minimal changes to the hot and cold dust amplitudes and correlations. Since there is no perceivable difference in the cold and hot dust amplitudes during the grid search we do not include those plots here. As a reminder, we chose to take the DIRBE fit values from \cite{CG02_05} for our fiducial values since the HFI have so little control of this dust component and as we expect the DIRBE bands to have better control of these SED parameters. 

\section{Thermal dust map characterization}
\label{app:dustCharacterization}
\begin{figure*}[h]
    \centering
    \includegraphics[width=0.9\linewidth]{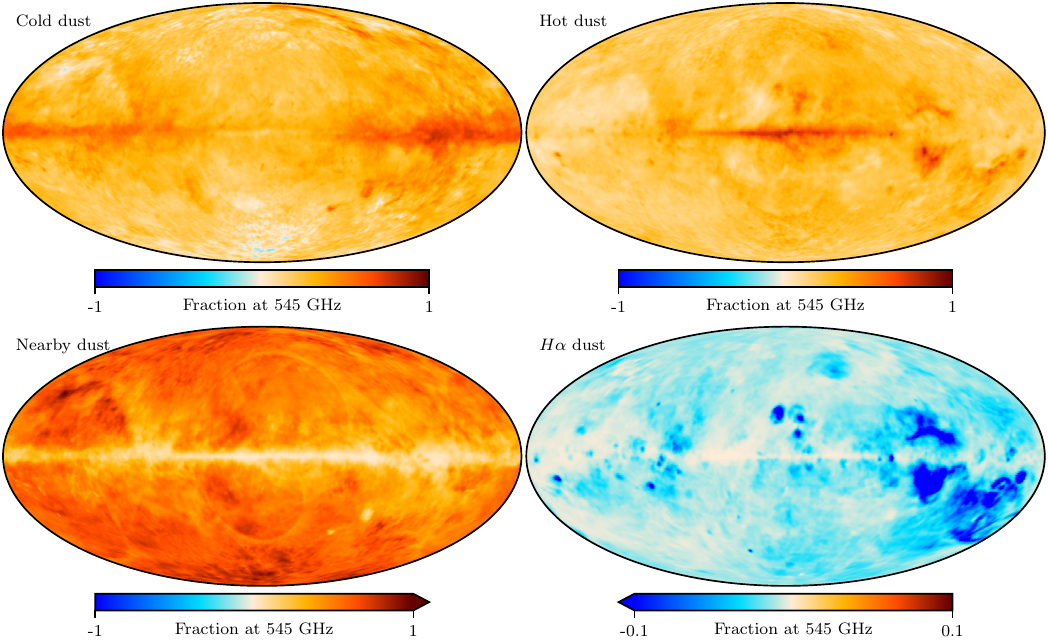}
    \caption{Average dust from each component contributing to the total dust signal. Note that the \Ha\ scale bar is a factor of ten lower.}
    \label{fig:dustFrac}
\end{figure*}

We compare the fraction of each dust component contributing to the total in \cref{fig:dustFrac}. As the \Ha\ dust is a dust extinction, with negative amplitude, it contributes a negative fraction, and it is the smallest contributor to the total. The nearby dust dominates outside the plane of the Galaxy (which is to be expected, since the nearer a dust cloud the greater its extent on the sky, and similarly further dust appears closer to the Galactic plane due to this geometry). The hot dust is clustered near the Galactic center and shows some anti-correlation with the \Ha\ dust. Finally, the cold dust is dominant in the Galactic plane further from the Galactic center. This further cements the notion that this updated dust model is more physical motivated than previous models, tracing known physical regions within the Milky Way Galaxy. 

\end{document}